\documentclass[a4paper,12pt]{article}
\RequirePackage{ifpdf}
\usepackage{comment}
\usepackage{soul}
\usepackage{jheppub}
\usepackage{esvect}
\usepackage{youngtab}
\usepackage{bbold}
\usepackage{float}
\usepackage{cancel}
\usepackage{braket}
\usepackage{tabularx}
\usepackage{mathtools, mathrsfs, hyperref,  amssymb, fancyhdr}
\usepackage{amsmath, graphics, setspace}
\usepackage{mathtools}
\usepackage{array}
\usepackage[table, dvipsnames]{xcolor}
\allowdisplaybreaks

\usepackage{tikz}
\usetikzlibrary{shapes,arrows,cd,chains,decorations.markings,decorations.pathmorphing,calc,positioning,patterns}

\tikzset{
	->-/.style args={#1rotate#2}{decoration={markings, mark=at position #1 with {\arrow[scale=1.5,rotate = #2 ]{stealth}}}, postaction={decorate}}
}
\tikzset{
	-r-/.style args={#1rotate#2}{decoration={markings, mark=at position #1 with {\arrow[scale=1,rotate = #2 ]{>}}}, postaction={decorate}}
}
\newcommand{\mathsym}[1]{{}}
\newcommand{\unicode}[1]{{}}

\newcolumntype{L}{>{$}l<{$}}
\definecolor{darkblue}{rgb}{0,0.1,.5}
\definecolor{darkred}{rgb}{.8,.1,0}
\definecolor{babyblueeyes}{rgb}{0.63, 0.79, 0.95}

\begin{document}
	\title{Looking for the \ensuremath{G_2} Higgs Branch of 4D Rank 1 SCFTs
 }
\author[1,3]{Md. Abhishek}
\author[2,3]{, Sachin Grover}
\author[2,3]{, Dileep P. Jatkar}
\author[4]{, and Kajal Singh}

\affiliation[1]{The Institute of Mathematical Sciences, IV Cross Road, CIT Campus, Taramani, \\ Chennai, India 600113. }

\affiliation[2]{Harish-Chandra Research Institute, Chhatnag Road, Jhunsi, Allahabad, India 211019. }

\affiliation[3]{Homi Bhabha National Institute, Training School Complex, Anushakti Nagar, Mumbai, India 400085. }

\affiliation[4]{Department of Mathematical Sciences, University of Liverpool, Liverpool, L69 7ZL, \\ United Kingdom.}

     \emailAdd{mdabhishek@imsc.res.in}
        \emailAdd{sachingrover@hri.res.in}
        \emailAdd{dileep@hri.res.in}
        \emailAdd{kajal.singh@liverpool.ac.uk}

	\abstract{The Schur index of the Higgs branch of 4-dimensional $\mathcal{N}=2$ SCFTs is related to the spectrum of non-unitary 2-dimensional CFTs. The rank 1 case has been shown to lead to the non-unitary CFTs with Deligne-Cvitanovic (DC) exceptional sequence of Lie groups. We show that a subsequence $(A_0, A_{\frac{1}{2}}, A_1, A_2, D_4)$ within the non-unitary sequence is related to a subsequence in the Mathur-Mukhi-Sen (MMS) sequence of unitary theories. We show that 2D non-unitary $G_2$ theory is related to unitary $E_6$ theory, and using this result along with the Galois conjugation, we propose that the $G_2$ Higgs branch is a sub-branch of the $E_6$ Higgs branch.}

 \makeatletter
\gdef\@fpheader{}
\makeatother
	
\maketitle
	
	
\section{Introduction}
There is an intriguing relation between the spectrum of the Schur operators in 4-dimensional (4D) $\mathcal{N}=2$ superconformal field
theories (SCFTs) and 2-dimensional (2D) non-unitary conformal field theories (CFTs) \cite{Beem:2013sza}. This relation has been explored in more detail in subsequent works \cite{Beem:2013sza,Beem:2014rza,Beem:2017ooy,Beem:2019tfp,Buican:2014qla,Buican:2015ina,Buican:2017rya,Buican:2019evc,Buican:2019huq,Buican:2021axn} (for a recent review and a comprehensive reference list, see \cite{Lemos:2020pqv}). In \cite{Beem:2013sza}, it was shown that the non-unitary 2D CFTs with current algebra symmetry belonging to affine versions of the Deligne-Cvitanovi\'c (DC) \cite{deligne1996serie,deligne1996serie2,Cvitanovic:2008zz} sequence of exceptional Lie groups with negative levels are related to certain Higgs branches of rank 1 theories in 4D, which saturate the unitarity bounds \cite{Lemos:2015orc, Beem:2017ooy}. This includes Argyres-Douglas (AD) theories with $A_0$, $A_1$, and $A_2$ flavour symmetry,\footnote{$A_0$ is the trivial symmetry.} the $SU(2)$ gauge theory with four flavours under $D_4$, and the Minahan–Nemeschansky theories for $E_6$, $E_7$, and $E_8$ symmetry. More precisely, \cite{Beem:2013sza} showed that the Schur index, which counts the Schur operators in $\mathcal{N}=2$ 4D SCFT, is given by the vacuum character of the 2D non-unitary DC series CFT at an appropriate negative level.  This relation opened up new exploration in 4D conformal bootstrap by utilising the 2D conformal bootstrap \cite{Kaidi:2022sng}.

{The DC series consists of $A_1$, $A_2$, $G_2$, $D_4$, $F_4$, $E_6$, $E_7$, and $E_8$ Lie algebras, which are the commutant subgroups of the $E_8$ Lie group.} Interestingly, Mathur, Mukhi, and Sen (MMS) \cite{Mathur:1988gt,Mathur:1988na,Mathur:1988rx} also stumbled upon the same series while addressing an entirely different problem. MMS were interested in classifying 2D CFTs with a fixed number of characters. In particular, MMS were attempting the classification of 2-character CFTs, and in that attempt, they found the DC series as well as a couple of additional CFTs with central charges $c=-22/5$ and $c=38/5$.  While the $c=-22/5$ theory was identified with the Lee-Yang edge singularity, $c=38/5$ was an enigmatic one at that time until it was found to possess an intermediate vertex operator algebra \cite{LANDSBERG2006143,Lee:2023owa}. The DC algebra appeared as an affine algebra at Ka\v c-Moody level 1 in the MMS set-up, and this entire set consisted of 2-character theories with effective central charge $0\leq c_{eff}\leq 8$. We will refer to this unitary series with $0\leq c_{eff}\leq 8$ as the MMS series and the same series for negative Ka\v c-Moody levels as the DC series with a tacit understanding that it is a series of non-unitary theories.

Returning to the results of \cite{Beem:2013sza}, the appearance of the DC series in the 4D rank 1 theories led to the following puzzle. In the study of the Higgs branch of Seiberg-Witten theory coupled to a certain number of hypermultiplets, it was known that we could get Higgs branches with global symmetry group $A_0$, $A_1$, $A_2$, $D_4$, $E_6$, $E_7$, and $E_8$. 
{It is evident that the groups $G_2$ and $F_4$ of the DC series are missing from this list. The $F_4$ theory is found to be anomalous in \cite{Shimizu:2017kzs}.} One of the aims of this paper is to find the $G_2$ among the Higgs branches of rank one theories. As mentioned earlier, the 4D-2D relation is such that the unitary 4D theory is related to non-unitary 2D CFT with the central charge relation $c_{2D} = -12 c_{4D}$ and $k_{2D} = - k_{4D}/2$.  Our strategy is to \textit{unitarise} these non-unitary theories. Unitarisation is a procedure in which one describes the set of highest weight characters in terms of a tentative theory with effective central charge $c_{eff} = c-24 h_{min}$, where $h_{min}$ is the minimum conformal dimension of the highest weight primary in the original theory \cite{Mathur:1988gt,Mathur:1988na, Grover:2022jrc}. If the original theory is unitary, then it automatically has the identity as minimum conformal dimension primary with $h_{min}=0$, but in non-unitary theories considered here, $h_{min}<0$. Using this unitarisation map, we transform a sequence of non-unitary theories into the sequence of tentative unitary theories.\footnote{Note that the theory obtained from a general non-unitary theory by the unitarisation map will not necessarily be a unitary theory although we have positive central charge and conformal dimensions.} There is an interesting shift in the rank of corresponding affine algebras under the unitarisation map. We discuss this in section \ref{unflavoured_correspondence}, where we show how a subsequence of the DC series gets mapped to a subsequence of the MMS series. We then show one isolated example ($G_2$ to $E_6$) from DC to MMS and the $F_4$ case, which maps to a theory with quasi-character.  

While the unflavoured characters are easy to relate under the unitarisation map, a more refined information can be extracted by turning on all the fugacities corresponding to the Cartan generators of the underlying affine symmetry.  We study the flavoured correspondence in section \ref{g2-e6-section} where we first consider the $A_1$ to $A_2$ relation, which is a part of the DC to MMS sequence. We show how the flavoured characters of $A_1$ theory can be obtained by restricting the fugacities of $A_2$ theory. There are multiple ways of doing this, but the purpose of this example is to show that even after turning all the fugacities on, the relation generated by the unitarisation map continues to hold albeit with restricted fugacities. We then examine the isolated case of relating $G_2$ at level $-5/3$ ($G_{2,-5/3}$) to $E_6$ at level 1 ($E_{6,1}$).  We first show that at the central charge $c=6$ besides having $E_{6,1}$, we also have $G_{2,3}$, which is a theory with six characters. By suitably choosing the basis we block diagonalise and pick up the two-character subsector of $G_{2,3}$ which maps to the $E_{6,1}$ characters.  Turning on flavours corresponds to 2 fugacities in the $G_2$ case and 6 in the $E_6$. One can get flavoured correspondence between $E_{6,1}$ and $G_{2,3}$ by judiciously choosing the fugacities of the $E_{6,1}$ theory. We then use the Galois conjugation method to provide another hint of the existence of the $G_2$ Higgs branch. The Galois conjugation relates commutant pairs within $E_{8,1}$ theory, which means $A_{2,1}$ and $E_{6,1}$ are Galois conjugates but these two theories are related to $A_{1,-4/3}$ and $G_{2,-5/3}$ respectively by the unitarisation map. We then use this flavoured correspondence to provide evidence of $G_2$ Higgs branch of rank one theories in 4D. 
We show that the modular data of $A_{1,-4/3}$ 
can be mapped by Galois conjugation to that of the $G_{2,-5/3}$ theory.\footnote{The modular data mapping by Galois conjugation is between the \textit{even sector} of the chiral algebras. We will describe the even sector and its relevance in the next section.} The relation between the $A_{1,-4/3}$ and $G_{2,-5/3}$ chiral algebras points towards the existence of the $G_2$ Higgs branch in 4D. 
We conclude this section by discussing line defects and defect Schur indices of the $G_2$ SCFT using the DC-MMS unitarisation map. In section \ref{flavouredDC}, we present the flavoured correspondence for the remaining DC-MMS series, and in section \ref{conclusions}, we briefly summarise our results and speculate about future directions.

    



\section{The unflavoured correspondence}\label{unflavoured_correspondence}

We begin with the explicit relationships between the non-unitary DC exceptional series CFTs and the corresponding unitary MMS series CFTs. A rational level $k_{2D}=t/u$, with $t\in\mathbb{Z}$ and $u\in\mathbb{Z}_{>0}$ coprime, is called Ka\v c-Wakimoto \textit{admissible} if $k_{2D}+h^\vee\geq h^\vee/u$ \cite{Kac:1988qc}. Generically, the non-unitary current algebra CFTs with an admissible fractional level are in general supposed to be logarithmic CFTs ($\log$ CFTs) based on results for lower rank current algebras \cite{Ridout:2008nh, Creutzig:2012sd, Creutzig:2013hma, Creutzig:2013yca}. The full category of modules, which is explicitly known for the lower rank current algebras $A_1$ and $A_2$, is a relaxed weight category including the indecomposable modules. The full category of modules of an admissible fractional level CFT has a subcategory $\mathcal{O}$ of the highest weight modules, which is closed under the modular transformations, but the Verlinde fusion coefficients are not positive definite. In addition, category $\mathcal{O}$ of a fractional level CFT is a non semi-simple module category \cite{Kawasetsu:2019att, Kawasetsu:2021qls}. Thus, the fractional level current algebra CFTs in the DC series are really $\log$ CFTs, except for the trivial current algebra $A_0$, which is a non-unitary minimal model.\footnote{We will not consider the current algebra at negative integer non-admissible level current algebras $D_4$, $E_6$, $E_7$ and $E_8$ occurring in DC series in this work.}

We have observed that the unflavoured characters of the subcategory $\mathcal{O}$ of highest weight modules of the DC series CFTs are the characters of the rational CFTs (RCFTs) with current algebra at level 1 appearing in the MMS series if we swap the role of the characters. The character of a highest weight module with current algebra $\hat{\mathfrak{g}}$ of rank $r$ in representation $\mathcal{R}$ is defined as,
\begin{equation}\label{char_def}
    \chi_{\mathcal{R}}(\tau, x_1,x_2\cdots, x_r)=Tr_{\mathcal{R}}q^{L_0-c/24}x_1^{H_1} x_2^{H_2}\cdots x_r^{H_r}\, ,
\end{equation}
where the modular parameter belongs to the upper-half complex plane $\tau\in \mathbb{H}$ and the fugacities $x_i\in U(1)$, with $1\leq i\leq r$. The nome is $q=e^{2\pi i \tau}$, and $H_i$'s are the Cartans of the finite algebra $\mathfrak{g}$.
The unflavoured limit of character is defined by turning off all flavour fugacities, i.e., $x_i\to 1$ for all $1\leq i\leq r$. The character \eqref{char_def} is a component of a vector that transforms as a vector valued modular function under the $SL(2,\mathbb{Z})$ modular transformations. The highest weight modules in the fractional level current algebra DC series CFTs as well as the unitary MMS series RCFTs have a 2-component unflavoured character vector. It is important to distinguish between the unflavoured characters and the flavoured characters since two or more modules can have the same unflavoured character whereas the flavoured characters are always in a one-to-one correspondence with the modules. Although the highest weight characters in a unitary CFTs have a well-defined unflavoured limit, the same is not true for the non-vacuum representations in the fractional level non-unitary CFTs.\footnote{We will only consider the highest weight representations in this paper unless otherwise noted.} The non-vacuum representations of the fractional level CFTs have a flavoured character with singularities at the unflavoured limit $(x_1,x_2,\cdots, x_r)=(1,1,\cdots,1)$. A well-defined finite unflavoured limit of the characters can only be obtained if we take a linear combination of characters with the same modular $T$-transformation which gets rid of the singularity in the unflavoured limit. This linear combination of characters, which has a finite unflavoured limit along with the vacuum character, constitutes the \textit{even sector} characters.\footnote{The nomenclature follows \cite{Mukhi:1989bp}.} Since the unflavoured vacuum character of the chiral algebra associated to a 4D $\mathcal{N}=2$ SCFT is the unflavoured Schur index, the Schur index itself is understood as a modular object \cite{Beem:2017ooy}.\footnote{The modularity of the flavoured Schur index can be exploited to predict important properties of its MLDE \cite{Pan2023, Pan:2024bne}.} The unflavoured Schur index then transforms to the even sector characters under modular $S$-transformations, which is the reason that we are interested in the even sector characters. However, in general, the unflavoured characters in the even sector are not always \textit{admissible}, i.e., although the characters are well-defined vector valued modular forms that satisfy the integrality of the coefficients of the $q$-series, the coefficients might not always satisfy positivity \cite{Grover:2022jrc}. Such characters that satisfy integrality but not positivity of the coefficients in their $q$-expansions are called \textit{quasi-characters}. We even find a quasi-character in the character vector of $F_{4,-5/2}$ in the DC series. We will discuss some speculations based on the appearance of the quasi-character in $F_{4,-5/2}$ CFT in the conclusions. The characters in the even sector satisfy a second order untwisted Modular Linear Differential Equation (MLDE). The MLDE for the DC exceptional series of non-unitary VOA is \cite{Kaneko:2013uga, Arakawa:2016hkg, Beem:2017ooy},
\begin{equation}\label{DCMLDE}
  \left(D_q^{(2)}-\frac{(h^{\vee}+1)(h^{\vee}-1)}{144}E_4(q)\right)\chi(q)=0,
\end{equation}
where $D_q^{(2)}$ is the second order modular differential operator and
$h^\vee$ is the dual Coxeter number specifying the level $k_{2D}=-h^{\vee}/6-1$ and the central charge $c_{2D}=-2(h^\vee+1)$. The function $E_4(q)$ is the normalised Eisenstein series of weight $4$. The two independent solutions of the eq(\ref{DCMLDE}) are,\footnote{Interestingly, if the dual Coxeter number $h^{\vee}=6n,\, n\in\mathbb{Z}_{>0}$, the corresponding VOA becomes Ka\v c-Wakimoto non-admissible and the modular differential equation (\ref{DCMLDE}) has a logarithmic solution. The logarithmic solutions were studied in \cite{Kaneko:2013uga}.}
\begin{eqnarray}\label{non_uni_ch}
\chi_0(q) & = & K(q)^{\frac{1+h^{\vee}}{12}}{}_2 F_1\left(\frac{1+h^{\vee}}{12},\,\frac{5+h^{\vee}}{12},\,1+\frac{h^{\vee}}{6},\,K(q)\right)\, ,\nonumber\\
\chi_1(q) & = & K(q)^{\frac{1-h^{\vee}}{12}}{}_2 F_1\left(\frac{1-h^{\vee}}{12},\,\frac{5-h^{\vee}}{12},\,1-\frac{h^{\vee}}{6},\,K(q)\right)\, ,
\end{eqnarray}
where $_2F_1$ are the usual hypergeometric functions. The rescaled inverse $K(q)$ of the Klein-invariant $j(q)$ is given in terms of the Eisenstein series $E_4$ and $E_6$,
\begin{equation}
    K(q)=\frac{1728}{j(q)}=\frac{E_4(q)^3-E_6(q)^2}{E_4(q)^3}\, ,
\end{equation}
where $E_4$ and $E_6$ are defined as,
\begin{align}
E_4(\tau) &:= 1+240 \sum_{n=1}^{\infty}\frac{n^3q^n}{1-q^n},\nonumber\\
E_6(\tau) &:= 1-504 \sum_{n=1}^{\infty}\frac{n^5q^n}{1-q^n}.
\end{align}
After normalisation, the $\chi_0(q)$ solution reproduces the vacuum character. The second solution $\chi_1(q)$ corresponds to the non-vacuum even character. Both characters $\chi_0(q)$ and $\chi_1(q)$ are the components of the even sector. The decomposition of the even character in terms of admissible highest weight representations can be easily found from the $S$-modular transformation matrix of the flavoured characters. 

%
%

To obtain a unitary CFT associated with the characters in \eqref{non_uni_ch}, we relabel the highest weight representations such that the character with the lowest conformal dimension is identified with the vacuum character of the tentative unitary theory. We refer to this procedure as \textit{unitarisation}. 
Importantly, unitarisation does not guarantee the existence of a unitary CFT in general. However, unitarisation of the even sector characters of the DC series CFTs produces the characters of the MMS series RCFTs, with the exception of the last map\footnote{The $F_{4,-5/2}$ has 2 independent characters in the unflavoured limit. The non-identity character is a quasi-character, which along with the vacuum character is identified with the two component character vector in class $D_4$ with the number of zeroes $\ell=0$ and $n=1$ mod 2 (type II) in the classification of \cite{Chandra:2018pjq}.} in \eqref{non-unitary to unitary series}. As a consequence of this result, we find the following relations through the unitarisation map,
    \begin{align}\label{non-unitary to unitary series}
	A_{0,-6/5} &\mapsto A_{\frac{1}{2},1}\, ,\nonumber\\
	A_{\frac{1}{2},-5/4} & \mapsto A_{1,1} \, ,\nonumber\\
	A_{1,-4/3} & \mapsto  A_{2,1}\, ,\nonumber\\
	A_{2,-3/2} & \mapsto D_{4,1}\, ,\nonumber\\
	G_{2,-5/3} & \mapsto E_{6,1} \, ,\nonumber\\
    F_{4,-5/2} & \mapsto F_{4,4} .
    \end{align}
This is the \textit{unflavoured correspondence} between the DC series CFTs and MMS series CFTs. For the remaining DC series chiral algebras, the non-vacuum conformal dimension is a negative integer. Upon unitarisation, the unitary theory should have two unflavoured characters with integer separation between their conformal dimensions. 
The integer separation between the conformal dimensions implies the order-2 MLDE would have a logarithmic solution as the non-vacuum character \cite{Arakawa:2016hkg, Beem:2017ooy}. Moreover, since the modular properties and the corresponding even sectors of the remaining DC exceptional groups, $D_4$, $E_6$, $E_7$, and $E_8$, are not well defined due to their Ka\v c-Wakimoto non-admissible negative integer levels and because their unitary counterpart does not belong to the MMS series, we restrict the unitarisation procedure to the chiral algebras of eqn \eqref{non-unitary to unitary series}. Later, we will also introduce discrete flavour fugacities in the unflavoured correspondence \eqref{non-unitary to unitary series}, whose details can be found in sections \ref{g2-e6-section} and \ref{flavouredDC}. In this section, our objective is to establish the unflavoured correspondence \eqref{non-unitary to unitary series}. We will write the first few terms of the $q$-expansion to demonstrate the unflavoured correspondence in each case of the series \eqref{non-unitary to unitary series}. The correspondence between $A_{0,-6/5}$ theory\cite{Beem:2013sza,Beem:2017ooy} (${\it aka}$ (2,5) minimal model) and $A_{\frac{1}{2},1}$ Intermediate VOA (IVOA) is well studied in literature \cite{Mathur:1988gt,Mathur:1988rx,Chandra:2018pjq}.\footnote{In \cite{Kawasetsu_2013}, the unitary theory with central charge $c=2/5$ and conformal dimension $h=1/5$ was identified with the IVOA with $A_{\frac{1}{2}}$.} The unitarisation map between the $A_{1,-4/3}$ CFT and the $A_{2,1}$ CFT has been studied in \cite{Mukhi:1989bp}, and the unitarisation map between the $A_{2,-3/2}$ CFT and the $D_{4,1}$ CFT was found in \cite{Buican:2017rya}.

In this section, we start by reviewing the well known example of this unflavoured correspondence between $A_{0,-6/5}$ and $A_{\frac{1}{2},1}$ theory; thereafter, we give a Coulumb gas representation of $A_{\frac{1}{2},-5/4}$, which establishes the second unitarisation map \eqref{non-unitary to unitary series} from $A_{\frac{1}{2},-5/4}$ CFT to $A_{1,1}$ CFT. We also establish the unflavoured correspondence between $G_{2,-5/3}$ and $E_{6,1}$ through unitarisation. Finally, we establish the last remaining relation in \eqref{non-unitary to unitary series} which is between the vacuum character of $F_{4,-5/2}$ CFT and the non-vacuum character of the highest weight representation $[1;0,0,1,1]$\footnote{The highest weight representation in terms of Dynkin labels of the affine algebra $\hat{\mathfrak{g}}$ of rank $r$ at Ka\v c-Moody level $k_{2D}$ is, $[\lambda_0;\lambda_1,\lambda_2,\cdots,\lambda_r]$ with $k_{2D}=\sum_{i=0}^r\lambda_i a_i^\vee$, where $a_i^\vee$ are the co-marks of the respective Dynkin nodes and $a_0^\vee=1$ \cite{DiFrancesco:1997nk}.} with classical dimension $1053$ of the current algebra $F_{4,4}$. This correspondence is limited since the non-vacuum character of $F_{4,-5/2}$ is a quasi-character.



\subsection{\ensuremath{A_{0,-6/5}} and \ensuremath{A_{\frac{1}{2},1}}}
The current algebra $A_{0,-6/5}$ CFT, also well known as Lee-Yang edge singularity CFT \cite{Kawasetsu_2013}, is realised as the $(2,5)$ minimal model with central charge $c=-22/5$.  The minimal model can be described in terms of two primary fields with conformal dimensions $h=0$, associated with the vacuum and the non-vacuum with $h=-1/5$. The characters of this CFT are,\footnote{We will denote the unflavoured characters with their conformal dimensions $\chi_h$ in this section.}
\begin{align}\label{a0-chars}
    \chi_0^{A_{0,-6/5}}&=q^{11/60}(1+q^2+q^3+q^4+\textit{O}(q^5))\, ,\nonumber\\
    \chi_{-1/5}^{A_{0,-6/5}}&=q^{-1/60}(1+q+q^2+q^3+2q^4+\textit{O}(q^5))\, .
\end{align}
The vacuum character $\chi_0^{A_{0,-6/5}}$ has also been shown to be the Schur index of the 4D $(A_1,A_2)$ Argyres-Douglas(AD) theory \cite{Cecotti:2010fi}. In \cite{Cecotti:2010fi}, it was shown that the Schur index of 4D $(A_1,A_{2n})$ should match with the vacuum character of the non-unitary $(2,2n+3)$ minimal model, and our $A_{0,-6/5}$ theory falls in this class. The corresponding unitary theory is obtained after unitarisation with effective central charge $c_{eff}=c-24h=\frac{2}{5}$, and corresponds to the $A_{\frac{1}{2},1}$ CFT \cite{Mathur:1988gt,Mathur:1988na,Mathur:1988rx,Chandra:2018pjq}. The unflavoured characters of this CFT are,
\begin{align}\label{a0-chars-exchanged}
    \chi_{0}^{A_{\frac{1}{2},1}}&=q^{-1/60}(1+q+q^2+q^3+2q^4+\textit{O}(q^5))\, ,\nonumber\\
    \chi_{1/5}^{A_{\frac{1}{2},1}}&=q^{11/60}(1+q^2+q^3+q^4+\textit{O}(q^5))\, .
\end{align}
The algebra has dim$(A_{1/2})=1$ which can be known from the $\textit{O}(q)$ coefficient and a `dual Coxeter number' $h^\vee=3/2$\cite{2cf0314cf8fd4509a1f3b973a384ae12}. This is a particularly interesting CFT since the current algebra does not belong to Ka\v c's classification (see for example \cite{kac_1990}).  One way to construct the $A_{1/2}$ theory at level 1 is by constructing a coset theory $E_8/E_{7\frac{1}{2}}$ where $E_{7\frac{1}{2}}$ CFT is identified as an \textit{intermediate vertex operator algebra} or IVOA \cite{Kawasetsu_2013, Marrani:2015nta, Lee:2023owa}.


\subsection{\ensuremath{A_{\frac{1}{2},-5/4}} and \ensuremath{A_{1,1}} }
In this subsection, our main result is a Coulomb gas representation of $A_{\frac{1}{2},-5/4}$, which is related to the free field representation of $A_{1,1}$. This will substantiate the second correspondence in the series \eqref{non-unitary to unitary series}.

The algebra $A_{1/2}$, although not a part of the original DC series \cite{deligne1996serie,Cvitanovic:2008zz}, fits nicely in between the trivial algebras $A_0$ and $A_1$ in the DC series when assigned a formal dual Coxeter number $h^\vee=3/2$ \cite{2cf0314cf8fd4509a1f3b973a384ae12}. The value $h^\vee=3/2$ when substituted in Deligne's dimension formula gives the dimension 1 in between the dimensions 0 and 3 for $A_0$ and $A_1$, respectively. From the 4-dimensional supersymmetric gauge theory point of view, the unitarity bounds of \cite{Beem:2013sza, Lemos:2015orc, Beem:2017ooy} are also saturated by the 2-dimensional chiral algebra $A_{\frac{1}{2},-5/4}$ like the rest of the members of the DC series.

Consider the $c=-5$ CFT with the current algebra $A_{\frac{1}{2},-5/4}$. The theory has two highest weight representations, with conformal dimensions $h=0$ and $h=-1/4$. Using eq.\eqref{non_uni_ch}, the characters of the $A_{\frac{1}{2},-5/4}$ CFT are,
\begin{align}\label{a1/2-chars}
    \chi_{0}^{A_{\frac{1}{2},-5/4}}&=q^{5/24}(1+q+3q^2+4q^3+7q^4+\textit{O}(q^5))\, ,\nonumber\\
    \chi_{-1/4}^{A_{\frac{1}{2},-5/4}}&=q^{-1/24}(1+3q+4q^2+13q^3+19q^4+\textit{O}(q^5))\, .
\end{align}
Under unitarisation it is easy to check that the characters are of the $A_{1,1}$ theory with the effective unitary central charge $c=1$ and conformal dimensions $h=\{0,\frac{1}{4}\}$. The unflavoured characters of $A_{1,1}$ are following,
\begin{align}\label{a1/2-chars-exchanged}
    \chi_{0}^{A_{1,1}}&=q^{-1/24}(1+3q+4q^2+13q^3+19q^4+\textit{O}(q^5))\, ,\nonumber\\
    \chi_{1/4}^{A_{1,1}}&=q^{5/24}(1+q+3q^2+4q^3+7q^4+\textit{O}(q^5))\, .
\end{align}
\subsubsection*{Coulomb gas representation of \ensuremath{A_{\frac{1}{2},-5/4}}}

In the Coulumb gas representation of $A_{\frac{1}{2},-5/4}$, we start with a chiral boson $\phi$ coupled with the scalar curvature $R$ with coupling $\gamma=i\sqrt{2}\alpha_0$, where $\alpha_0$ is the background charge. In this construction, the central charge of the scalar field theory with the dilaton coupling is (see \cite{DiFrancesco:1997nk} for further details),
\begin{equation}
    c = 1-24\alpha_0^2\, .
\end{equation}
To construct the representation of $A_{\frac{1}{2},-5/4}$ theory with $c=-5$, we require the background charge to be $\alpha_0 = 1/2$. In this Coulomb gas model, we have a dimension $h=1$, a primary $\partial \phi$, and an infinite number of Virasoro primaries, which are the vertex operators, $V_{\alpha} = \exp(i\sqrt{2}\alpha\phi)$, with screening charge $\alpha$ and conformal dimension,
\begin{equation}\label{coulomb conformal dimension}
    h= \alpha^2 - 2\alpha_0\alpha.
\end{equation}
The non-trivial Virasoro primary with the conformal dimension $h=-1/4$ is the vertex operator $V_{1/2} = \exp(i\phi/\sqrt{2})$ with a screening charge $\alpha = 1/2$. A closer look at the unflavoured character of the primary $V_{1/2}$ in eq.\eqref{a1/2-chars} tells us that there are three operators at grade one. Apart from the operator $L_{-1}V_{1/2}$, we need to find two more operators at grade one, i.e., with conformal dimension $h=3/4$. After solving eq.\eqref{coulomb conformal dimension} with $h=3/4$, we get the solution set $\alpha = 3/2, -1/2$. We now have two extra operators,
\begin{equation}
    V_{3/2} = \exp(3i\phi/\sqrt{2}),\ \quad
V_{-1/2} = \exp(-i\phi/\sqrt{2}),
\end{equation}
 which are contributing to the first grade of the module of the $V_{1/2}$ primary. We change the vertex operator “momentum" from $\alpha\to\alpha-1/2$ to obtain the free field representation of the unitary $A_{1,1}$ CFT. Note that the Cartan $H=\partial\phi$ remains unchanged while the vertex operators are shifted to,
\begin{equation}
  V_{3/2} = \exp(i(3/2)\sqrt{2}\phi)  \to E^{+}=\exp(i\sqrt{2}\phi)\, ,
\end{equation}
and,
\begin{equation}
V_{-1/2} = \exp(-i(1/2)\sqrt{2}\phi) \to E^{-}=\exp(-i\sqrt{2}\phi)\, .
\end{equation}
These operators $E^{\pm}$ are precisely the raising and lowering generators of $A_{1,1}$ current algebra with the Cartan $H=\partial\phi$.

\subsection{\ensuremath{G_{2,-5/3}} and \ensuremath{E_{6,1}}}

Our next candidate of the series \eqref{non-unitary to unitary series} is the correspondence $G_{2,-5/3} \mapsto E_{6,1}$. The non unitary current algebra $G_{2,-5/3}$ CFT has central charge $c=-10$ and has two Ka\v c-Wakimoto admissible highest weight representations \cite{Kac:1988qc} with conformal dimension of $h=-\frac{2}{3}$ \cite{axtell2010vertex}. Using eq.(\ref{non_uni_ch}) the unflavoured characters of this theory are given by,
\begin{align}\label{g2-e6-unflavoured}
\chi^{G_{2,-5/3}}_0 & =  q^{5/12}\left(1+14 q+92 q^{2}+456 q^{3}+1848 q^{4}+\textit{O}(q^5)\right),\nonumber\\
\chi^{G_{2,-5/3}}_{-2/3} & =  {q^{-1/4}}\left(1+78 q+729 q^{2}+4382 q^{3}+19917 q^{4}+\textit{O}(q^5)\right).
\end{align}
Under the unitarisation and the following unflavoured character flip,
\begin{align}
   &\chi^{G_{2,-5/3}}_0(q,1,1)\sim\chi^{E_{6,1}}_{2/3}(q)\, , \\
    &\chi^{G_{2,-5/3}}_{[-1/3;0,-2/3]}(q,1,1)-\chi^{G_{2,-5/3}}_{[0;1,-4/3]}(q,1,1)=\chi^{G_{2,-5/3}}_{-2/3}(q)\sim\chi^{E_{6,1}}_{0}(q)\, .
\end{align}
In the later sections, this unflavoured correspondence will be discussed in more detail by turning on the flavour fugacities and studying the modular properties under the Galois conjugation.

\subsection{Non unitary \ensuremath{F_{4,-5/2}} theory and quasi characters}
The last candidate of the series of unflavoured correspondence \eqref{non-unitary to unitary series} is the non-unitary current algebra $F_4$ CFT at level $k_{2D}=-\frac{5}{2}$ with $c_{2D}=-20$. The conformal dimensions of the Ka\v c-Wakimoto admissible highest weights are $h_{min}=\{0,-\frac{3}{2}\}$ and have the following unflavoured characters from \eqref{non_uni_ch},
\begin{eqnarray}\label{f4-fractional-level-chars}
\chi^{F_{4,-5/2}}_0 & = & q^{5/6}\left(1+52 q+1106 q^{2}+14808 q^{3}+ \textit{O}[q^4]\right),\nonumber\\
\chi^{F_{4,-5/2}}_{-3/2} & = & {q^{-2/3}}\left(1-272 q-34696 q^{2}-1058368 q^{3}-\textit{O}[q^4]\right).
\end{eqnarray}
The character $\chi^{F_{4,-5/2}}_{-3/2}$ is a quasi character with a single positive coefficient and countably infinite negative coefficients. Under unitarisation the role of the characters flips, and we obtain the type II quasi-character in the $D_4$ series with $\ell=0$ in the classification of \cite{Chandra:2018pjq}. We identify the vacuum characters in eqn \eqref{f4-fractional-level-chars} with the unflavoured characters of $F_{4,4}$ CFT,
\begin{align}
    &\chi^{F_{4,-5/2}}_0(q)\sim \chi^{F_{4,4}}_{[1;0,0,1,1]}(q)\, \nonumber\\
    &\chi^{F_{4,-5/2}}_{-3/2}(q)\sim \chi^{F_{4,4}}_{[4;0,0,0,0]}(q)-\chi^{F_{4,4}}_{[2;0,0,0,2]}(q)-\chi^{F_{4,4}}_{[0;0,1,0,1]}(q)\, .
\end{align}
Note that the `vacuum character' in the unitary side is a quasi-character. 

\section{\ensuremath{G_2-E_6} correspondence and 4D SCFT}\label{g2-e6-section}
We will now elaborate on the correspondence between the highest weight characters of the $G_{2}$ chiral algebra at level $k_{2D}=-5/3$ and the RCFT with chiral algebra $E_6$ at level 1. Using this correspondence between the non-unitary $G_2$ CFT and the $E_6$ WZW model, we propose that the SCFT with $G_2$ flavour symmetry can be constructed from F-theory compactification at constant coupling $\tau=e^{\pi i/3}$ where the $E_6$ point is located\cite{Sen:1996vd, Dasgupta:1996ij}. We further reveal another surprising connection between the modular data of the highest weight representations of $A_1$ current algebra at level $-4/3$ and $G_2$ current algebra at level $-5/3$ via Galois conjugation. This powerful relation predicts observables in the $G_2$ 4-dimensional SCFT from observables of $(A_1,A_3)$ Argyres-Douglas (AD) theory.

\subsection{Flavouring the \texorpdfstring{$A_1-A_2$}{A1-A2} correspondence}\label{flavouring_a1_a2}

Firstly, let us present a toy example where we will relate the non-vacuum character of $A_2$ current algebra at level 1 with the vacuum character of $A_{1}$ current algebra at level $-4/3$. We will follow the same algorithm to produce a flavoured $G_2-E_6$ correspondence.

We will show the flavoured correspondence between the non-vacuum character of the RCFT with current algebra $A_{2,1}$ and the vacuum character of the current algebra $A_{1,-4/3}$. The relation between the characters of these two algebras is not straightforward since they have different ranks. Thus, to find a relation, we need to turn off a fugacity in $A_{2,1}$ current algebra. The remaining fugacity of $A_{2,1}$ is exactly the fugacity of $A_{1,-4/3}$ algebra. However, it is instructive as well as relevant for the higher rank algebras to relate the fugacities via a conformal embedding \cite{Buican:2019huq}. We demonstrate the procedure for $A_1/A_2$ correspondence below. 

\begin{table}
\centering
\begin{tabularx}{0.8\textwidth} { 
  | >{\centering\arraybackslash}X 
  | >{\centering\arraybackslash}X 
  | >{\centering\arraybackslash}X | }
 \hline
 $A_{2,1}$ representations & character labels ($\chi_{\mathrm{dim}\mathcal{R}})$ & conformal weight ($h$) \\
\hline
$[1;0,0]$  & $\chi_\mathbf{1}$  & $0$ \\
\hline
$[0;1,0]$  & $\chi_\mathbf{3}$  & $1/3$ \\
\hline
$[0;0,1]$  & $\chi_\mathbf{\bar 3}$  & $1/3$ \\
\hline
\end{tabularx}
\caption{The highest weight representations of $A_2$ at level 1 are tabulated with the character labels $\chi_{\mathrm{dim}\mathcal{R}}$ in each representation $\mathcal{R}$ and the conformal weight $h(\mathcal{R})$.}
\label{table: A_2_1_reps}
\end{table}

The highest weight representations of $A_{2,1}$ algebra are given in table \ref{table: A_2_1_reps}. The relevant conformal embedding is $A_{1,4}\subset A_{2,1}$. The highest weight representations of $A_{1,4}$ have $A_1$ weights $2s$ and conformal weights,\footnote{For many general integrable representations of a current algebra, the branching rules in terms of finite algebra can be obtained from the book of \cite{kass1990affine}. Branching rules can in general be computed using procedures on \textit{SageMath}. For example, check this \href{https://doc.sagemath.org/html/en/reference/combinat/sage/combinat/root_system/branching_rules.html}{link}. }
\begin{equation}
    (2s+1,h)=\left(1,0\right)\, , \left(2,\frac{1}{8}\right)\, ,\left(3,\frac{1}{3}\right)\, ,\left(4,\frac{5}{8}\right)\, ,\left(5,1\right)\, .
\end{equation}
There exists a non-anomalous $\mathbb{Z}_2$ symmetry, which we use to orbifold the $A_{1,4}$ CFT. The orbifold CFT $A_{1,4}/\mathbb{Z}_2$ has the partition function,
\begin{equation}
    \mathcal{Z}_{A_{1,4}/\mathbb{Z}_2}=|\chi_{\bf 1}+\chi_{\bf 5}|^2+2|\chi_{\bf 3}|^2\, ,
\end{equation}
where $\chi_{\mathbf{2s+1}}$ is the (unflavoured) character of representation with spin $s$. This is a $D$-type modular invariant in the classification of the partition function of $A_1$ CFTs \cite{Cappelli:1987xt}. In fact this is the partition function of $A_{2,1}$ CFT as well, where we make the identification,
\begin{equation}
    \chi_{\bf 1}^{A_{2,1}}=\chi_{\bf 1}^{A_{1,4}}+\chi_{\bf 5}^{A_{1,4}}\, ,\quad \chi_{\bf 3}^{A_{2,1}}=\chi_{\mathbf{\bar 3}}^{A_{2,1}}=\chi_{\bf 3}^{A_{1,4}}\, .
\end{equation}
Note that the number of fugacities in the identification is not the same. The unnormalised flavoured non-vacuum characters $\chi_{\bf 3}^{A_{2,1}}$ and $\chi_{\mathbf{\bar 3}}^{A_{2,1}}$ in terms of the characters $\mathrm{ch}^{A_2}_{\mathrm{dim}\mathcal{R}}(x_1,x_2)$ of finite Lie algebra $A_{2}$ are,\footnote{See appendix \ref{a2 basics} details of the finite characters of $A_2$ algebra.}
\begin{align}\label{a_2_1_flavoured_chars}
    \chi_{\bf 3}^{A_{2,1}}(q,x_1,x_2)&=\mathrm{ch}^{A_2}_{\bf 3}+q(\mathrm{ch}^{A_2}_{\bf 3}+\mathrm{ch}^{A_2}_{\mathbf{\bar 6}})+q^2(2\mathrm{ch}^{A_2}_{\bf 3}+\mathrm{ch}^{A_2}_{\mathbf{\bar 6}}+\mathrm{ch}^{A_2}_{\bf 15})+\textit{O}(q^3)\, ,\nonumber\\
    \chi_{\mathbf{\bar 3}}^{A_{2,1}}(q,x_1,x_2)&=\mathrm{ch}^{A_2}_{\mathbf{\bar 3}}+q(\mathrm{ch}^{A_2}_{\mathbf{\bar 3}}+\mathrm{ch}^{A_2}_{\bf 6})+q^2(2\mathrm{ch}^{A_2}_{\mathbf{\bar 3}}+\mathrm{ch}^{A_2}_{\bf 6}+\mathrm{ch}^{A_2}_{\mathbf{\bar 15}})+\textit{O}(q^3)\, ,
\end{align}
 where we have suppressed the flavour fugacity variables $x_1$ and $x_2$ in the argument on the right-hand side. The characters in \eqref{a_2_1_flavoured_chars} match exactly with the unnormalised non-vacuum character of $A_{1,4}$,
\begin{equation}
    \chi_{\bf 3}^{A_{1,4}}(q,x)=\mathrm{ch}^{A_1}_{\bf 3}+q(1+\mathrm{ch}^{A_1}_{\bf 3}+\mathrm{ch}^{A_1}_{\bf 5})+q^2(1+3\mathrm{ch}^{A_1}_{\bf 3}+ 2\mathrm{ch}^{A_1}_{\bf 5}+\mathrm{ch}^{A_1}_{\bf 7})+\textit{O}(q^3)\, ,
\end{equation}
if we turn off the flavour fugacity $x_2$ and recognise $x_1=x$,\footnote{Of course there are other relations as well, for example turning off $x_1$ and recognising $x_2=x$.} i.e,
\begin{equation}\label{a_1_4_flavoured_char}
    \chi_{\bf 3}^{A_{2,1}}(q,x,1)=\chi_{\mathbf{\bar 3}}^{A_{2,1}}(q,x,1)=\chi_{\bf 3}^{A_{1,4}}(q,x)\, .
\end{equation}
\begin{table}
\centering
\begin{tabularx}{0.8\textwidth} { 
  | >{\centering\arraybackslash}X 
  | >{\centering\arraybackslash}X 
  | >{\centering\arraybackslash}X | }
 \hline
 $A_{1,-4/3}$  highest weights & character labels & conformal weight ($h$) \\
\hline
$[-4/3;0]$  & $\chi_0$  & $0$ \\
\hline
$[-2/3;-2/3]$  & $\chi_1$  & $-1/3$ \\
\hline
$[0;-4/3]$  & $\chi_2$  & $-1/3$ \\
\hline
\end{tabularx}
\caption{The highest weight representations of $A_{1}$ at level $-4/3$ are tabulated with the character labels $\chi_i$ and the conformal weight $h(\mathcal{R})$.}
\label{table: A_1_43_reps}
\end{table}
The unflavoured vacuum character of $A_{1,-4/3}$ algebra is the non-vacuum unflavoured character $\chi_{\bf 3}^{A_{1,4}}(q,1)$. To find other values of flavour fugacity where this correspondence holds, we introduce discrete flavour fugacity, which are the solutions of the equation \cite{Buican:2019huq},
\begin{equation}\label{A_1_discrete_fugacities}
    \chi_{\bf 3}^{A_{1,4}}(q,x^i)=\mathcal{N}^{A_1}(x)\cdot\chi_{0}^{A_{1,-4/3}}(q,x^i)\, ,
\end{equation}
where the vacuum character has $A_1$ the representations given in table \ref{table: A_1-43_reps}.

\begin{table}
\centering
\begin{tabular}{ |p{2cm}|p{8cm}|  }
\hline
grade & $A_ 1$ representations and multiplicities \\
\hline
0 & {\bf 1} \\
\hline
1 &  {\bf 3} \\
\hline
2 & {\bf 1}, {\bf 3}, {\bf 5}\\
\hline
3 & {\bf 1}, 2({\bf 3}), {\bf 5}, {\bf 7}\\
\hline
4 & 2({\bf 1}), 3({\bf 3}), 3({\bf 5}), {\bf 7}, {\bf 9}\\
\hline
\end{tabular}
\caption{The multiplicities of each $A_1$ representation which occurs at the first five grades in the vacuum representation $[-4/3;0]$ of $A_{1,-4/3}$ current algebra.}
\label{table: A_1-43_reps}
\end{table}

Although we can try to solve this equation analytically using the $\Theta(q,x)$ functions of \cite{Kac:1988qc}, we will demonstrate the validity of our algorithm using only flavoured $q$-series, which can be easily used for higher rank algebras. We will use this technique when we flavour the $G_2-E_6$ correspondence as well in the next subsection. Starting with the $q$-expansion in \eqref{a_1_4_flavoured_char} and the $q$-expansion of the vacuum character $\chi_{0}^{A_{1,-4/3}}$, we obtain the following results order by order of $q$:
\begin{itemize}
    \item At $\textit{O}(q^0)$, since the vacuum character $\chi_{0}^{A_{1,-4/3}}$ has unit coefficient and the left-hand side of \eqref{A_1_discrete_fugacities} has the finite character, we have $\mathcal{N}^{A_1}(x)=\mathrm{ch}^{A_1}_{\bf 3}$.
    \item At $\textit{O}(q)$ and $\textit{O}(q^2)$, the character on both sides of \eqref{A_1_discrete_fugacities} match for every value of fugacity. 
    \item Only at $\textit{O}(q^3)$, we find that the fourth roots of unity are the only fugacity solutions.  Since the character combination $1+\mathrm{ch}^{A_1}_{\bf 3}+\mathrm{ch}^{A_1}_{\bf 7}$ can always be found at $\textit{O}(q^3)$ or higher, this implies a simultaneous solution of the polynomial equations at every order (which is a polynomial of a increasing degree with increasing order) must belong to the solution set of the equation at $\textit{O}(q^3)$. This result matches with the analytic calculation of \cite{Buican:2019huq}.
\end{itemize}

The vacuum character of the $A_{1,-4/3}$ current algebra is the Schur index of the $(A_1,A_3)$ Argyres-Douglas (AD) theory. Thus, the Schur index of $(A_1,A_3)$ AD theory with discrete values of fugacity can be obtained from the non-vacuum character of $A_{2,1}$ chiral algebra after we switch off one of the fugacities and recognise the remaining fugacity with $A_1$ fugacity but only at discrete values, i.e. fourth root of unity. 

\subsection{Flavouring \texorpdfstring{$G_2-E_6$}{G2/E6} correspondence}\label{Flavouring G2/E6 correspondence}
We enrich the unflavoured correspondence between $G_{2,-5/3}$ and $E_{6,1}$ chiral algebra by introducing discrete flavours into the correspondence. In particular, we will switch off some flavour fugacities in $E_{6,1}$ non-vacuum character and relate the remaining fugacities with the fugacities of corresponding $G_{2,3}$ character, which is embedded inside $E_{6,1}$.

The unflavoured correspondence between $G_{2,-5/3}$ and $E_{6,1}$ is noted in eqn \eqref{g2-e6-unflavoured}. To introduce fugacity in the unflavoured correspondence, we utilise the conformal embedding of $G_{2,3}$ inside $E_{6,1}$ denoted by $(G_2)_3\subset(E_6)_1$. The primaries of $G_2$ at level 3 theory with the highest weight representations and corresponding conformal dimensions are written in table \ref{G_2_3_reps}.

\begin{table}
\centering
\begin{tabularx}{0.8\textwidth} { 
  | >{\centering\arraybackslash}X 
  | >{\centering\arraybackslash}X 
  | >{\centering\arraybackslash}X | }
 \hline
 $(G_2)_3$ representations & character labels ($\chi_{\mathrm{dim}\mathcal{R}})$ & conformal weight ($h$) \\
\hline
\rowcolor{babyblueeyes}
$[3;0,0]$  & $\chi_\mathbf{1}$  & $0$ \\
\hline
$[2;0,1]$  & $\chi_\mathbf{7}$  & $2/7$ \\
\hline
\rowcolor{babyblueeyes}
$[1;0,2]$  & $\chi_\mathbf{27}$  & $2/3$ \\
\hline
$[1;1,0]$  & $\chi_\mathbf{14}$  & $4/7$ \\
\hline
$[0;0,3]$  & $\chi_\mathbf{77}$  & $8/7$ \\
\hline
\rowcolor{babyblueeyes}
$[0;1,1]$  & $\chi_\mathbf{64}$  & $1$ \\
\hline
\end{tabularx}
\caption{The highest weight representations of $G_2$ at level 3 are tabulated with the character labels $\chi_{\mathrm{dim}\mathcal{R}}$ in each representation $\mathcal{R}$ and conformal weight $h(\mathcal{R})$. The representations marked in blue are the ones which are part of the orbifold $G_{2,3}/\mathbb{Z}_2$ CFT.}
\label{G_2_3_reps}
\end{table}
The modular $S$-matrix of $(G_2)_3$ is \cite{Coquereaux:2010we},
\begin{equation}
    S=\frac{1}{\sqrt{21}}\begin{pmatrix}
\sqrt{\frac{5-\sqrt{21}}{2}} & \sqrt{3} & \sqrt{7} & \sqrt{3} & \sqrt{3} & \sqrt{\frac{5+\sqrt{21}}{2}}\\
\sqrt{3} & \sqrt{3}u_3 & 0 & \sqrt{3}u_2 & \sqrt{3}u_1 & -\sqrt{3} \\
\sqrt{7} & 0 & -\sqrt{7} & 0 & 0 & \sqrt{7} \\
\sqrt{3} & \sqrt{3}u_2 & 0 & \sqrt{3}u_1 & \sqrt{3}u_3 & -\sqrt{3} \\
\sqrt{3} & \sqrt{3}u_1 & 0 & \sqrt{3}u_3 & \sqrt{3}u_2 & -\sqrt{3} \\
\sqrt{\frac{5+\sqrt{21}}{2}} & -\sqrt{3}& \sqrt{7} & -\sqrt{3} & -\sqrt{3} & \sqrt{\frac{5-\sqrt{21}}{2}}
\end{pmatrix},
\end{equation}
where $u_1<u_2<u_3\in\mathbb{R}$ are roots of the polynomial $u^3-u^2-2u+1$, (this polynomial appears due to $\mathbb{Z}_3$ fusion rules as seen in the $MTC$ classification \cite{rowell2009classification}). 

The $G_2$ CFT at level 3 also has the central charge $c_{2D}=6$, same as the central charge of $E_6$ CFT at level 1. The $G_2$ CFT at level 3 has a subsector which transforms into itself under modular transformations, and by this virtue, it also has a closed fusion algebra. We utilise the fact that the representations  $[3;0,0]$ and $[0;1,1]$ have integer-separated conformal dimensions to write the even and odd linear combination of these two characters. Thus, the vector-valued modular form transforms under modular $S$-transformation as two disjoint vector valued modular forms with a block-diagonal modular $S$-matrix and the same modular $T$-matrix as before,

\begin{equation}\label{g2_3_s_matrix}
    \begin{pmatrix}
        \chi_{\mathbf{1}}+\chi_{\mathbf{64}}\\
        \chi_{\mathbf{27}}\\
        \chi_{\mathbf{1}}-\chi_{\mathbf{64}}\\
        \chi_{\mathbf{7}}\\
        \chi_{\mathbf{14}}\\
        \chi_{\mathbf{77}}
    \end{pmatrix}(-1/\tau)=\begin{pmatrix}
\frac{1}{\sqrt{3}} & \frac{2}{\sqrt{3}} & 0 & 0 & 0& 0\\
\frac{1}{\sqrt{3}} & -\frac{1}{\sqrt{3}} & 0 & 0 & 0 & 0 \\
0 & 0 & -\frac{1}{\sqrt{7}} & \frac{2}{\sqrt{7}} & \frac{2}{\sqrt{7}} & \frac{2}{\sqrt{7}} \\
0 & 0 & \frac{1}{\sqrt{7}} & \frac{u_3}{\sqrt{7}} & \frac{u_2}{\sqrt{7}} & \frac{u_1}{\sqrt{7}} \\
0 & 0 & \frac{1}{\sqrt{7}} & \frac{u_2}{\sqrt{7}} & \frac{u_1}{\sqrt{7}} & \frac{u_3}{\sqrt{7}} \\
0 & 0 & \frac{1}{\sqrt{7}} & \frac{u_1}{\sqrt{7}} & \frac{u_3}{\sqrt{7}} & \frac{u_2}{\sqrt{7}}
\end{pmatrix}\begin{pmatrix}
        \chi_{\mathbf{1}}+\chi_{\mathbf{64}}\\
        \chi_{\mathbf{27}}\\
        \chi_{\mathbf{1}}-\chi_{\mathbf{64}}\\
        \chi_{\mathbf{7}}\\
        \chi_{\mathbf{14}}\\
        \chi_{\mathbf{77}}
    \end{pmatrix}(\tau)\, .
\end{equation}

We identify the unflavoured characters,\footnote{We have added the names $E_{6,1}$ and $G_{2,3}$ of the current algebra in the superscript to distinguish between their respective characters.}\textsuperscript{,}\footnote{The second set contains a quasi-character $(\chi_{(0,0)}-\chi_{(1,1)})(1,1,q)$ as can be easily checked. In the second set since there is no good vacuum character,i.e., the unflavoured character has a $q$-expansion of the kind $\chi_0(q)=q^{-c_{eff}/24}(1+a_1q+a_2q^2+\cdots)$ with $a_i\in\mathbb{Z}_{\geq 0}$ we do not analyse it further.}
\begin{align}\label{g2-e6-char-rel}
    \chi^{E_{6,1}}_1 =\chi^{G_{2,3}}_\mathbf{1}+\chi^{G_{2,3}}_\mathbf{64}, &\qquad \chi^{E_{6,1}}_\mathbf{27}=\chi^{E_{6,1}}_\mathbf{\overline{27}}=\chi^{G_{2,3}}_\mathbf{27}\, ,
    \end{align}
    and identify the top left block -diagonal $S$ matrix with the modular $S$-matrix of $E_{6,1}$ CFT,
\begin{equation}\label{e6-modular-s}
    S=\frac{1}{\sqrt{3}}\begin{pmatrix}
1 & 1 & 1\\
1 & \omega & \omega^2 \\
1 & \omega^2 & \omega
\end{pmatrix}; \qquad \omega=e^{\frac{2\pi i}{3}}\, ,
\end{equation}
and the correct modular $T$-matrix
\begin{align}\label{e6-modular-t}
    T=diag\{e^{-i\pi/2},e^{5i\pi/6},e^{5i\pi/6}\}\, .
\end{align}
The identification in the previous equations \eqref{g2-e6-char-rel}, \eqref{e6-modular-s}, \eqref{e6-modular-t} relate the two partition functions as,
\begin{align}
    \mathcal{Z}[E_{6,1}]&=|\chi^{E_{6,1}}_\mathbf{1}|^2+|\chi^{E_{6,1}}_\mathbf{27}|^2+|\chi^{E_{6,1}}_\mathbf{\overline{27}}|^2\, ,\nonumber\\
    \mathcal{Z}[G_{2,3}]&=|\chi^{G_{2,3}}_\mathbf{1}+\chi^{G_{2,3}}_\mathbf{64}|^2+2|\chi^{G_{2,3}}_\mathbf{27}|^2\, .
\end{align}
This identification matches the old  observations in \cite{Christe:1988vc,Coquereaux:2010we}. The theory obtained by choosing only the representations $[3;0,0], [0;1,1]$ and $[1;0,2]$ of the $G_{2,3}$ algebra is the $\mathbb{Z}_2$ orbifold theory $G_{2,3}/\mathbb{Z}_2$, where $[1;0,2]$ occurs twice since it is the fixed point. 

We make use of the conformal embedding $G_{2,3}$ inside $E_{6,1}$ to discover the solutions of the fugacity matching conditions. 
\begin{center}
\begin{tabularx}{0.8\textwidth} { 
  | >{\centering\arraybackslash}X 
  | >{\centering\arraybackslash}X 
  | >{\centering\arraybackslash}X | }
 \hline
 $(G_2)_3$ highest weights & character labels ($\chi_{\mathrm{dim}\mathcal{R}}$) & conformal weight ($h$) \\
\hline
$[3;0,0]$  & $\chi_{\bf 1}$  & $0$ \\
\hline
$[1;0,2]$  & $\chi_{\bf 27}$  & $\frac{2}{3}$ \\
\hline
$[0;1,1]$  & $\chi_{\bf 64}$  & $1$ \\
\hline
\end{tabularx}
\end{center} 
\begin{center}
\begin{tabularx}{0.8\textwidth} { 
  | >{\centering\arraybackslash}X 
  | >{\centering\arraybackslash}X 
  | >{\centering\arraybackslash}X | }
 \hline
 $(E_6)_1$ highest weights & character labels ($\chi_{\mathrm{dim}\mathcal{R}}$)& conformal weight ($h$) \\
\hline
$[1;0,0,0,0,0,0]$  & $\chi_{\bf{1}}$  & $0$ \\
\hline
$[0;1,0,0,0,0,0]$  & $\chi_{\bf 27}$  & $\frac{2}{3}$ \\
\hline
$[0;0,0,0,0,1,0]$  & $\chi_{\bf{\overline{27}}}$  & $\frac{2}{3}$ \\
\hline
\end{tabularx}
\end{center} 

To introduce flavour fugacities in the relations \eqref{g2-e6-char-rel}, we first write the non-vacuum flavoured characters, which will be related to the vacuum character of $G_{2,-5/3}$,
\begin{table}
\centering
\begin{tabular}{ |p{2cm}|p{8cm}|  }
\hline
level & $G_ 2$ representations and multiplicities \\
\hline
0 & {\bf 27} \\
\hline
1 &  {\bf 7}, {\bf 14}, {\bf 27}, {\bf 64}, {\bf 77}, {\bf 189} \\
\hline
2 & {\bf 1}, 2({\bf 7}), 2({\bf 14}), 5({\bf 27}), 4({\bf 64}), 3({\bf 77}),  2({\bf 77$'$}), 2({\bf 182}),  3({\bf 189}), {\bf 286}, {\bf 448}\\
\hline
3 & {\bf 1}, 7({\bf 7}), 8({\bf 14}), 11({\bf 27}), 13({\bf 64}), 12({\bf 77}),\\ &  5({\bf 77$'$}), 6({\bf 182}),  12({\bf 189}), {\bf 273}, 5({\bf 286}),\\ & 2({\bf 378}), 5({\bf 448}), {\bf 729}, {\bf 924}\\
\hline
4 & 5({\bf 1}), 16({\bf 7}), 18({\bf 14}), 33({\bf 27}), 37({\bf 64}), 32({\bf 77}),  19({\bf 77$'$}), 23({\bf 182}),  35({\bf 189}), 3({\bf 273}), 17({\bf 286}), 8({\bf 378}), 19({\bf 448}), 8({\bf 729}), 2({\bf 714}), 5({\bf 924}), {\bf 896}, {\bf 1547}\\
\hline
\end{tabular}
\caption{The multiplicities of each $G_2$ representation which occurs at the first five grades in the non-vacuum representation $[1;0,2]$ of $G_{2,3}$ current algebra.}
\label{table: g2-level-3-reps}
\end{table} 
\begin{equation}\label{e6-level1-non-vac}
    \chi^{E_{6,1}}_\mathbf{27}=\mathrm{ch}^{E_6}_\mathbf{27}+q(\mathrm{ch}^{E_6}_\mathbf{27}+\mathrm{ch}^{E_6}_\mathbf{351})+ q^2(2\,\mathrm{ch}^{E_6}_\mathbf{27}+\mathrm{ch}^{E_6}_\mathbf{351}+\mathrm{ch}^{E_6}_\mathbf{351'}+\mathrm{ch}^{E_6}_\mathbf{1728})+\textit{O}(q^3)\, , 
\end{equation}

\begin{table}
\centering
\begin{tabular}{ |p{2cm}|p{8cm}|  }
\hline
level & $E_ 6$ representations and multiplicities \\
\hline
0 & {\bf 27} \\
\hline
1 &  {\bf 27}, {\bf 351} \\
\hline
2 & 2({\bf 27}), {\bf 351}, {\bf 351$'$}, {\bf 1728}\\
\hline
3 & 3({\bf 27}), 3({\bf 351}), {\bf 351$'$}, 2({\bf 1728}), {\bf 7371}\\
\hline
4 & 6({\bf 27}), 5({\bf 351}), 3({\bf 351$'$}), 4({\bf 1728}), 2({\bf 7371}), {\bf 7722},  {\bf 17550}\\
\hline
\end{tabular}
\caption{The multiplicities of each $E_6$ representation which occurs at the first five grades in the non-vacuum representation $[0;1,0,0,0,0,0]$ of $E_{6,1}$ current algebra.}
\label{table: e6-level-1-reps}
\end{table} 
and,
\begin{align}\label{g2-level3-non-vacuum}
    \chi^{G_{2,3}}_\mathbf{27}&=\mathrm{ch}^{G_2}_\mathbf{27}+
    q(\mathrm{ch}^{G_2}_\mathbf{7}+\mathrm{ch}^{G_2}_\mathbf{14}+\mathrm{ch}^{G_2}_\mathbf{27}+\mathrm{ch}^{G_2}_\mathbf{64}+\mathrm{ch}^{G_2}_\mathbf{77}+\mathrm{ch}^{G_2}_\mathbf{189})+q^2(1+2\mathrm{ch}^{G_2}_\mathbf{7}\nonumber\\
    &\quad+2\mathrm{ch}^{G_2}_\mathbf{14}+5\mathrm{ch}^{G_2}_\mathbf{27}+4\mathrm{ch}^{G_2}_\mathbf{64}+3\mathrm{ch}^{G_2}_\mathbf{77}+2\mathrm{ch}^{G_2}_\mathbf{77'}+2\mathrm{ch}^{G_2}_\mathbf{182}+3\mathrm{ch}^{G_2}_\mathbf{189}\nonumber\\
    &\quad+\mathrm{ch}^{G_2}_\mathbf{286}+\mathrm{ch}^{G_2}_\mathbf{448})+\textit{O}(q^3)\, . 
\end{align}
 The characters of $G_2$ finite Lie algebra, which are used to write the characters of the representation $\bf 27$ of $G_{2,3}$ are functions of 2 fugacities $(y_1,y_2)$ which can be written in terms of the six flavour fugacities of $E_6$ $(x_1,x_2,\cdots,x_6)$ with the identification,
\begin{align}\label{e6-flavours-to-g2-flavours}
    x_1&=x_5=1\, ,\nonumber\\
    x_2&=y_1\, ,\quad x_4=y_2\nonumber\\
    x_3&=\frac{1}{y_1 y_2}\, ,\quad  x_6=\frac{1}{y_2}\, .  
\end{align}
Note that this is not a unique identification, with similar identifications generated by symmetry operations on the one given above. The details of these identifications can be obtained in the Appendix \ref{discrete_flavours}. 

We now have the ingredients to relate the non-vacuum character of $G_2$ CFT at level 3 to the vacuum character of $G_2$ CFT at level $-5/3$ at discrete flavour fugacity. We first note the admissible representations of $(G_2)_{-5/3}$ \cite{axtell2010vertex,Kawasetsu:2019att}.
\begin{table}
\centering
\begin{tabularx}{0.8\textwidth} { 
  | >{\centering\arraybackslash}X 
  | >{\centering\arraybackslash}X 
  | >{\centering\arraybackslash}X | }
 \hline
 $(G_2)_{-5/3}$ highest weights & character labels & conformal weight ($h$) \\
\hline
$[-\frac{5}{3};0,0]$  & $\chi_0$  & $0$ \\
\hline
$[-\frac{1}{3};0,-\frac{2}{3}]$  & $\chi_{1}$  & $-\frac{2}{3}$ \\
\hline
$[0;1,-\frac{4}{3}]$  & $\chi_2$  & $-\frac{2}{3}$ \\
\hline
\end{tabularx}
\label{table: non-unitary-g2-hw}
\end{table} 
The module corresponding to the highest weight $[-\frac{5}{3};0,0]$ is finite-dimensional, and the character is the vacuum character. The other two highest weight modules are infinite-dimensional, and their combination gives finite unrefined character $\Tilde{\chi}_{-2/3}$ with $x_1\to 1$ and $x_2\to 1$. To introduce flavour into the $G_2-E_6$ correspondence, we write the flavoured vacuum character of $G_{2,-5/3}$ CFT in terms of the characters of $G_2$ finite algebra using the Ka\v c-Wakimoto character formula (see appendix \ref{Details of vacuum character} for details). Not writing the 2-dimensional CFT normalisation $q^{-\pi i c_{2D}/12}$ such that the first term is unity,
\begin{align}\label{g2-frac-vacuum}
    q^{-5/12}\chi_{0}^{G_{2,-5/3}}&= 1+q\mathrm{ch}_{\mathbf{14}}+q^2(1+\mathrm{ch}_{\mathbf{14}}+\mathrm{ch}_{\mathbf{77'}})+q^3(1+2\mathrm{ch}_{\mathbf{14}}+\mathrm{ch}_{\mathbf{77}}+\mathrm{ch}_{\mathbf{77'}}+\mathrm{ch}_{\mathbf{273}})\nonumber\\
&\quad+q^4(2+3\mathrm{ch}_{\mathbf{14}}+\mathrm{ch}_{\mathbf{27}}+\mathrm{ch}_{\mathbf{77}}+3\mathrm{ch}_{\mathbf{77'}}+\mathrm{ch}_{\mathbf{273}}+\mathrm{ch}_{\mathbf{448}}+\mathrm{ch}_{\mathbf{748}})+\textit{O}(q^5).
\end{align}

\begin{table}
\centering
\begin{tabular}{ |p{2cm}|p{8cm}|  }
\hline
grade $[q^n]$ & $G_ 2$ representations and their multiplicities \\
\hline
0 & {\bf 1} \\
\hline
1 & {\bf 14} \\
\hline
2 & {\bf 1}, {\bf 14}, {\bf 77$'$}\\
\hline
3 & {\bf 1}, 2({\bf 14}), {\bf 77}, {\bf 77$'$}, {\bf 273}\\
\hline
4 & 2({\bf 1}), 3({\bf 14}), {\bf 27}, {\bf 77}, 3({\bf 77$'$}), {\bf 273}, {\bf 448}, {\bf 748}\\
\hline
\end{tabular}
\caption{The multiplicities of each $G_2$ representation which occurs at the first five grades in the vacuum representation $[-5/3;0,0]$ of $G_{2}$ at level $-5/3$ current algebra.}
\label{table: g2-m5/3-mutiplicity}
\end{table}

We denote by $\ket{\Omega}$ the primary state corresponding to the unique vacuum of the $G_{2,-5/3}$ theory with conformal dimension $h=0$. Since the vacuum state is non-degenerate, the $q^0$ coefficient is just the character of the singlet representation, $\text{ch}_{\bf 1}$. At grade $one$ we have total 14 states, corresponding to the affine currents $J_{-1}^a$\footnote{The corresponding Schur operator is the \textit{moment map} operator of the flavour-current multiplet $\hat{\mathcal{B}}_1$ transforming in the adjoint representation of the flavour group $G_2$ \cite{Beem:2013sza,Beem:2017ooy}. } acting on the state $\ket{\Omega}$,
\begin{align}
    \alpha_aJ_{-1}^a\ket{\Omega}\,,\qquad a=1,2,\cdots,14\,.
\end{align}
At grade $two$, we have the following states,
\begin{align}\label{grade 2}
    \left(\beta_{(ab)}J_{-1}^{(a}J_{-1}^{b)}+\gamma_cJ_{-2}^c\right)\ket{\Omega}\,.
\end{align}
%
The first term, symmetric under $(a, b)$, yields a total of $\bf{sym^2(adj)}$ states. The symmetric squared adjoint representation decomposes into the representation with twice the Dynkin label of the adjoint representation  $2(\bf{adj})$, and the Joseph ideal $\mathcal{I}_2$,
\begin{equation}
    \bf{sym}^2(adj)=\mathcal{I}_2\oplus 2(adj)\,,\qquad  \mathcal{I}_2=1\oplus\mathfrak{R}\,.
\end{equation}
For the group $G_2$, the representation $\bf{\mathfrak{R}=27}$ \cite{Cvitanovic:2008zz}. 
Among these states, there are null states corresponding to the representation $\bf{27}$ of $G_2$ \cite{Arakawa:2015jya}. The remaining states of the form $\gamma_cJ^c_{-2}\ket{\Omega}$ correspond to the representation $\mathbf{14}$,\footnote{In 4D, the presence of null states in the vacuum module of the 2D $G_2$ CFT implies that the contribution of Higgs chiral ring multiplet $\hat{\mathcal{B}}_2$ in representation $\mathfrak{R}=27$ channel of the OPE with $\hat{\mathcal{B}}_1$ vanishes. In fact, the contribution of $\hat{\mathcal{B}}_2$ to the OPE with $\hat{\mathcal{B}}_1$ in the entire \textit{Joseph ideal} $\mathcal{I}_2$ vanish \cite{Beem:2013sza,Beem:2017ooy}. 
The non-vanishing flavour singlet channel contribution only comes from the three point coupling, $\langle \hat{\mathcal{B}}_1\hat{\mathcal{B}}_1\hat{\mathcal{C}}_{0(0,0)}\rangle$, with the unique stress-tensor multiplet $\hat{\mathcal{C}}_{0(0,0)}$.}
\begin{align}\label{joseph_ideal_vac_char}
    [\bf{sym^2(14)]\oplus14=[1\oplus \cancel{27} \oplus 77'] \oplus 14}.
\end{align}
Similarly we can understand the appearance of certain representations at higher orders $O(q^n)$ in the vacuum character \eqref{g2-frac-vacuum}.

The unflavoured characters of the $G_{2,-5/3}$ are mapped to the unflavoured characters of the $G_{2,3}$ under unitarisation by utilising the unflavoured correspondence \eqref{g2-e6-unflavoured} and the identification from conformal embedding\eqref{g2-e6-char-rel}.\footnote{The conformal embedding of non-unitary $G_{2,-5/3}$ to non-unitary $A_{2,-5/3}$ current algebra CFT is studied in \cite{adamovic2011general}. } We guess that at discrete values of fugacities labelled by $(y_1^i,y_2^i)$ are the $i$-th solution of the equation,
 \begin{equation}\label{fugacity_match-ansatz_g2}
        \chi^{G_{2,3}}_{\bf 27}(q,y_1^i,y_2^i)=\mathcal{N}(y_1^i,y_2^i)\chi^{G_{2,-5/3}}_{0}(q,y_1^i,y_2^i)\, ,
    \end{equation}
where $\mathcal{N}(y_1^i,y_2^i)$ is a function of fugacities only. We have checked that our ansatz has solutions at roots of unity as expected from a similar analysis done in \cite{Buican:2019huq} for the simple $A_1$ current algebra. Although the eq. \eqref{fugacity_match-ansatz_g2} should hold at every order in $q$, we argue below that the set of solutions at lower order (for this case, we only need $q^2$) are the set of solutions for the equation \ref{fugacity_match-ansatz_g2} at every other order. We have also checked the validity of this claim to $q^5$ not just for $G_2$ but a few other members in the series \eqref{non-unitary to unitary series}. 

We first argue that the function $\mathcal{N}(y_1,y_2)=\mathrm{ch}_{\bf 27}(y_1,y_2)$ at all values of fugacities $y_1\, ,y_2\in u(1)$ due to the matching required at $\textit{O}(q^0)$. Since we have identified the polynomial $\mathcal{N}(y_1,y_2)$ all we need to do is solve a polynomial equation in $y_1$ and $y_2$ at each order in $q$ obtained from expanding the relation \eqref{fugacity_match-ansatz_g2}.
\begin{itemize}
    \item The $\textit{O}(q)$ polynomial equation is identically satisfied for each value of the variables $y_1$ and $y_2$.
    \item We find that at the next order $\textit{O}(q^2)$, the polynomial equation is non-trivial (see apendix \ref{non vac G2 3} for more detail). Solving the $\textit{O}(q^2)$ polynomial equation gives discrete flavour fugacity solutions at roots of unity, i.e., $(y_1,y_2)=(e^{2\pi i /n_1},e^{2\pi i /n_2})$, where $n_1\, ,n_2\in \{7,10,13\}$. The full set of solutions, of which 17 solutions determine the rest, is provided in the section \ref{explicit_fug_g2}.
    \item At the next few orders we checked explicitly, $\textit{O}(q^3)$ and $\textit{O}(q^4)$, we get the same set of solutions we obtained while solving at $\textit{O}(q^2)$. This is justified by the fact that on the left and the right hand side of the equation, the character combination $1+\mathrm{ch}^{G_2}_{\bf 14}+\mathrm{ch}^{G_2}_{\bf 77'}$ and the combination $1+2ch^{G_2}_{\bf 7}+2\mathrm{ch}^{G_2}_\mathbf{14}+5\mathrm{ch}^{G_2}_\mathbf{27}+4\mathrm{ch}^{G_2}_\mathbf{64}+3\mathrm{ch}^{G_2}_\mathbf{77}+2\mathrm{ch}^{G_2}_\mathbf{77'}+2\mathrm{ch}^{G_2}_\mathbf{182}+3\mathrm{ch}^{G_2}_\mathbf{189}+\mathrm{ch}^{G_2}_{\bf 286}+\mathrm{ch}^{G_2}_{\bf 448}$ respectively can always be found at $\textit{O}(q^3)$ or higher, which implies a simultaneous solution of the polynomial equations at every order (which is a polynomial of a increasing degree with increasing order) must belong to the solution set of the equation at $\textit{O}(q^3)$.  Consequently, we have the lowest number of solutions, which every other polynomial equation (at higher order in $q$) needs to satisfy. 
\end{itemize}

To summarise, we find 61 such roots of unity solutions, of which 17 are independent up to exchange and complex conjugation symmetries explained in equations \eqref{symmetry_1} and \eqref{symmetry_2} respectively. These independent solutions are the set of 7th, 10th, and 13th roots of unity, respectively. The explicit details are given in subsection \ref{explicit_fug_g2}.

\subsection{Galois conjugation}\label{galois_conjugation_section}

We now take a sharp turn and reveal a surprising connection between $A_{1,-4/3}$ and $G_{2,-5/3}$ chiral algebras. This connection can be explained using the following map,
\begin{equation}\label{figure-gal-uni}
	   \begin{tikzpicture}                             
		\draw[->] (-1,1) -- (1,1);
		\draw[->] (-1,-1) -- (1,-1);
		\draw[->] (-3,1) -- (-2.5,1);
		\draw[->] (3,-1) -- (2.5,-1);
		\draw[->] (3,1) -- (2.5,1);
		\draw[->] (-3,-1) -- (-2.5,-1);
		\draw (-3,1) -- (-3,-1);
		\draw (3,1) -- (3,-1);
		\node at (-1.8, 1) {$(A_1)_{-\frac{4}{3}}$};
		\node at (1.8, 1) {$(A_2)_{1}$};
		\node at (-1.8, -1) {$(G_2)_{-\frac{5}{3}}$};
		\node at (1.8, -1) {$(E_6)_{1}$};
	\node at (0,0) {Unitarisation};
	\node at (-5,0) {Galois conjugates};
	\node at (5,0) {Galois conjugates};	
	\end{tikzpicture}                    
\end{equation}
 
The fact that the commutant pairs $A_{2,1}$ and $E_{6,1}$ inside $E_{8,1}$ are Galois conjugates is known from previous results \cite{rowell2009classification}.\footnote{The commutant pairs inside $E_{8,1}$ are all Galois conjugates as well \cite{rowell2009classification}. This can be pushed to the other \textit{non-unitary counterparts}, but we are only interested in $G_{2,-5/3}$ category $\mathcal{O}$ in this section.}
The action of the Galois transformation on the modular data (the modular $S$ and $T$ matrices) is easier to describe than its full action on the MTC associated with an RCFT. \footnote{According to the conjecture that each RCFT is realised by a unique MTC. Although the inverse is not true since the pentagon and the hexagon equations can have non-modular solutions.} The quantum dimensions, which are ratios of elements of the modular $S$ matrices, belong to the cyclotomic field $\mathbb{Q}(\xi(n))$, which is a field of rationals multiplied by (up to) $n$-th roots of unity, which we denote by $\xi(n)=e^{2\pi i /n}$. The Galois transformations form a group called the \textit{Galois group} denoted by $\mathbb{Z}_{n}^{\times}$. The Galois action is naturally defined by the roots of unity,
\begin{equation}\label{galois-action-def}
\xi(n)\mapsto\xi(n)^p\, ,
\end{equation}
where $p\in\mathbb{Z}_{n}^{\times}$ while leaving the field of rationals intact. The Galois action on the modular $S$ and $T$ matrices can be calculated by the action \eqref{galois-action-def}. The details can be found in \cite{Coste:1993af, Harvey:2018rdc, DiFrancesco:1997nk}.\footnote{The Galois group $\mathbb{Z}_n^\times$ can also be calculated from a general formula in \cite{DiFrancesco:1997nk}, where $n=M(k+h^{\vee})$ and $M=N$ for SU(N), and 3 for $G_2$. But the group $\mathbb{Z}_n^\times$ might not be faithful.}

Let us explain the Galois action on the modular data of $A_{1,-4/3}$ current algebra. The 2-dimensional CFT with the spectrum generating algebra $(A_1)_{-4/3}$ has the 2-dimensional central charge $c_{2D}=-6$. In its full glory, it is a $\log$ CFT, but it has a sector of 3 highest weight modules in category $\mathcal{O}$ with conformal dimensions $h=\{0,-\frac{1}{3},-\frac{1}{3}\}$ \cite{Creutzig:2012sd, Creutzig:2013hma, Creutzig:2013yca}. The flavoured characters of the 3 modules are closed under the modular transformations $S$ and $T$ \cite{Mathur:1988gt, Dedushenko:2018bpp},
\begin{equation}
    S_{A_{1,-4/3}}=\frac{1}{\xi(12)+\xi(12)^{-1}}\begin{pmatrix}
\xi(12)^6 & 1 &\xi(12)^6\\
1 & \xi(12)^2 & \xi(12)^4 \\
\xi(12)^6 & \xi(12)^4 & \xi(12)^2
\end{pmatrix}\, ,
\end{equation}
\begin{align}
    T=\mathrm{diag}(1,\xi(12)^4,\xi(12)^4)\, ,
\end{align}
where $\xi(n):=e^{2\pi i/n}$, and we have ignored the RCFT normalisation $\varphi=e^{\pi i/2}$ in writing the modular $T$-transformation. The Galois group is just $\mathbb{Z}_3^{\times}$ instead of $\mathbb{Z}_{12}^{\times}$ since $T^3=\mathbb{1}$.\footnote{Here we have used the notion of an MTC conductor $N_0$ such that $T^{N_0}=\mathbb{1}$, borrowing the name from \cite{Buican:2019evc} which is similar to Bantay's conductor $N$ for an RCFT for which $(\varphi T)^N=\mathbb{1}$ \cite{Bantay_2003}.} The Galois group is similar to the unitary counterpart $A_{2,1}$ RCFT, which also has a $\mathbb{Z}_3^{\times}$ Galois group owing to the $\mathbb{Z}_3$ fusion algebra \cite{Buican:2021axn}. In fact, up to signs $A_{1,-4/3}$ highest weight representations have a $\mathbb{Z}_3$ Verlinde algebra (see, for example, \cite{Buican:2019huq}). The Galois group $\mathbb{Z}_3^{\times}$ has only one non-trivial element $2\in\mathbb{Z}_3^\times$, under which we get the complex conjugated modular data $(S^*,T^*)$. Interestingly, up to the RCFT normalisation, the $T^*$ is nothing but the $T$ matrix of $(G_2)_{-5/3}$ theory with conformal dimension $h=\{0,-\frac{2}{3},-\frac{2}{3}\}$ and central charge $c=-10$.
\begin{align}
    T_{A_{1,-4/3}}&=\mathrm{diag}(1,\xi(3),\xi(3))\, ,\\
    T_{G_{2,-5/3}}&=\mathrm{diag}(1,\xi(3)^2,\xi(3)^2)\, .
\end{align}

The modular-$S$ matrix $S_{A_{1,-4/3}}$ transforms under Galois transformation to modular-$S$ matrix of $G_{2,-5/3}$,
\begin{equation}\label{s-matrix-g2-53}
  S_{G_{2,-5/3}}=S_{A_{1,-4/3}}^{*}=\frac{1}{\xi(12)+\xi(12)^{-1}}\begin{pmatrix}
\xi(12)^6 & 1 &\xi(12)^6\\
1 & \xi(12)^4 & \xi(12)^2 \\
\xi(12)^6 & \xi(12)^2 & \xi(12)^4
\end{pmatrix}\, .
\end{equation}
The modular-$S$ matrix of $G_{2,-5/3}$ can be derived from unflavoured $G_2-E_6$ correspondence. The way to do it is to start with the $2\times 2$ block diagonal modular-$S$ matrix for the unflavoured characters of $G_{2,3}/\mathbb{Z}_2$ orbifold CFT or the $E_{6,1}$ RCFT \eqref{g2 basics}.\footnote{The modular-$S$ matrix of the unflavoured character is also called the \textit{reduced} $S$ matrix.} Since the unflavoured characters of $G_{2,-5/3}$ can be obtained by a character swap of the $E_{6,1}$ unflavoured characters, the reduced modular-$S$ matrix for $G_{2,-5/3}$ is,
\begin{equation}
    S_{G_{2,-5/3}}^{red}=\frac{1}{\sqrt{3}}\begin{pmatrix}
      -1 & 1\\
      2 & 1
    \end{pmatrix}
\end{equation}.
Then we can obtain the $3\times 3$ modular-$S$ matrix \eqref{s-matrix-g2-53} by writing a symmetric $S$-matrix following the procedure in \cite{Mathur:1988gt}. 

The Galois action shuffles the diagonal and off-diagonal entries in the lower $2\times 2$ block of the modular-$S$ matrix while remaining entries remain the same. It is the same Galois action that relates the $A_{2,1}$ and $E_{6,1}$ chiral algebras on the unitary side, as depicted in figure \eqref{figure-gal-uni}. The $G_{2,-5/3}$ modular-$S$ matrix is simply written as,
\begin{equation}\label{g2_-5_3_s-matrix}
  S_{G_{2,-5/3}}=-\frac{1}{\sqrt 3}\begin{pmatrix}
1 & -1 &1\\
-1 & \xi(3) & -\xi(3)^2 \\
1 & -\xi(3)^2 & \xi(3)
\end{pmatrix}\, ,
\end{equation}
from which we write the \textit{even} combination of highest weight characters that has a well-defined limit, i.e., $\chi_{0}^{G_{2,-5/3}}(q,1,1)$ and $\chi_{ 1}^{G_{2,-5/3}}(q,1,1)-\chi_{ 2}^{G_{2,-5/3}}(q,1,1)$ are well defined. Since Galois conjugation will preserve Verlinde fusion rules, the fusion algebra of $G_{2,-5/3}$ is the $\mathbb{Z}_3$ fusion algebra, up to a transformation $[\phi_1]\to -[\phi_1]$,
\begin{align}
    [\phi_1]\otimes [\phi_2]&=-[\phi_0]\, ,\nonumber\\
    [\phi_1]\otimes [\phi_1]&=[\phi_2]\, ,\nonumber\\
    [\phi_2]\otimes [\phi_2]&=-[\phi_1]\, ,   
\end{align}
where the modules associated with the vacuum are denoted by $[\phi_0]$, and the two modules $[\phi_1]$ and $[\phi_2]$ correspond to the representations $[-1/3;0,-2/3]$ and $[0;1,-4/3]$, respectively. The negative fusion rules are expected since we know that category $\mathcal{O}$ is not closed and we would need a relaxed highest weight category including the indecomposable modules to build the correct fusion rules for the $\log$ CFT \cite{Ridout:2014yfa}.


\subsection{\texorpdfstring{$G_2$}{G2} Higgs Branch from Flavoured Correspondence}

Until now, we have used a couple of relations to study non-unitary CFTs; one of them is the unitarisation map, and another is the Galois conjugation. The unitarisation map relates a non-unitary theory to a unitary theory, and in our case, we have established a relation between the Ka\v c-Wakimoto admissible levels in the DC series relevant to rank one superconformal theory in 4D, and the MMS series \eqref{non-unitary to unitary series}. The Galois conjugation, on the other hand, gives a relationship between commutant pairs of CFTs in 2D. If one of the theories within the commutant pairs exists, then the Galois conjugation guarantees the existence of the other theory in that pair. When we combine the unitarisation map with the Galois conjugation, we get a stronger result. The argument requires gathering results from all the steps mentioned above, and therefore, we will consider only the $G_2-E_6$ case here. Next, we will construct evidence in favour of the $G_2$ Higgs branch in 4D theory following the results for its chiral algebra.

Recall that the $G_2-E_6$ correspondence is an isolated example of the unitarisation map. The 2D theory, which describes the $G_2$ Higgs branch of the 4D superconformal theory as its associated variety, has $G_{2,-5/3}$ affine symmetry with a central charge $c=-10$. The Schur index of the $G_2$ SCFT, which is the vacuum character of the $G_{2,-5/3}$ chiral algebra \eqref{g2-frac-vacuum}, corresponds to the saturation of the unitarity bounds on the 4D $G_2$ SCFT. In other words, the Schur index tells us that the $\hat{\mathcal{B}}_2$ multiplet does not contribute to the OPE with $\hat{\mathcal{B}}_1$ in representation $\mathfrak{R}=27$ \eqref{joseph_ideal_vac_char}, which implies that the closure of the minimal nilpotent orbit of $G_2$ should be isomorphic to the Higgs branch of the $G_2$ SCFT \cite{Beem:2013sza, Beem:2017ooy, Beem:2019tfp}. The 2D $G_2$ theory under the unitarisation map goes over to $c=6$ CFT with $E_{6,1}$ symmetry when we look at it with all  flavour fugacities turned off. However, once we consider all flavour fugacities, then $E_{6,1}$ possesses four more fugacities compared to $G_{2,-5/3}$. We then notice that at $c=6$, we also have a CFT with $G_{2,3}$ symmetry. The flavour correspondence with identification given in eq.\eqref{e6-flavours-to-g2-flavours} relates the $E_{6,1}$ modules to the even sector of the $G_{2,3}$ modules. We thus see that the DC theory with $G_{2,-5/3}$ with all fugacities turned on is related to $G_{2,3}$ at $c=6$ through flavour correspondence between $E_{6,1}$ and $G_{2,3}$.

Before we get into the implications of 4D theory, let us apply the Galois conjugation to the $G_2-E_6$ case. In the unflavoured sector, $A_{1,-4/3}$ is mapped to $A_{2,1}$ using the unitarisation map and  $G_{2,-5/3}$ goes over to $E_{6,1}$. The Galois conjugation, on the other hand, is a relation between commutant pairs inside $E_{8,1}$. Therefore, $E_{6,1}$ is Galois conjugate to $A_{2,1}$ by virtue of being commutants in $E_{8,1}$. The Galois conjugation guarantees that if a CFT corresponding to one member of the commutant pair exists, then a CFT for the commutant pair also exists. If we run the unitarisation map in the reverse direction, then the commutant pairs $A_{2,1}$ and $E_{6,1}$ map to the category $\mathcal{O}$ of $A_{1,-4/3}$ and $G_{2,-5/3}$ theory, respectively. Since $A_{1,-4/3}$ theory is known to exist, the Galois conjugation relation guarantees the existence of $G_{2,-5/3}$ theory. Lifting the correspondence between the chiral algebras to their associated varieties has interesting implications for the Higgs branch with $G_2$ symmetry.

The arguments given above suggest that if we have to look for the Higgs branch in 4D superconformal theory with $G_2$ symmetry, then we should look at the theory with $E_6$ symmetry and tune the flavour fugacities as per the identification given in eq.\eqref{e6-flavours-to-g2-flavours}. If we take the F-theory approach to study the Higgs branches of rank one theories \cite{Sen:1996vd,Dasgupta:1996ij}, then it is immediately obvious that the underlying four manifolds is supposed to be K3, and it is known from the work of Kronheimer \cite{Kronheimer:1989zs} that the four manifolds with SU(2) holonomy have only ADE-type singularities. This clearly rules out a $G_2$ type singularity. However, this does not rule out the possibility of the $G_2$ Higgs branch being a subbranch of another Higgs branch. It is natural to look at the $E_6$ Higgs branch and restrict the fugacities as given in eq.\eqref{e6-flavours-to-g2-flavours} to find the $G_2$ Higgs branch. Notice that this is a subbranch of the $E_6$ Higgs branch because we have to set a couple of fugacities to 1, and restrict the remaining four fugacities so that they can be written in terms of two parameters.

It is interesting to note that $E_6$ symmetry does occur in the F-theory compactification to eight dimensions, which is a set-up for the $\mathcal{N}=2$ SU(2) gauge theory in four dimensions. More interestingly, the $E_6$ symmetry occurs with constant coupling \cite{Dasgupta:1996ij}. Unlike the $D_4$ symmetry, which can occur for any constant coupling \cite{Sen:1996vd}, the $E_6$ symmetry can occur only when the coupling is constant and is given by $\tau = \exp(i\pi/3)$ \cite{Dasgupta:1996ij}. We therefore claim that the Higgs branch with $G_2$ symmetry occurs at a constant coupling $\tau = \exp(i\pi/3)$. This branch was not located earlier simply because it is a subbranch of the $E_6$ branch and occurs at the same point in the coupling constant moduli space. 

Roughly the setup can be described as follows. Starting from $E_6$ singularity, where we have coincident D-branes, we separate the branes. The distance $d$ between these branes makes strings stretched between these separated D-branes massive with mass $m \propto d/\sqrt{\alpha'}$. Massive strings are stretched between the D-branes which break the $E_6$ symmetry spontaneously. In fact, the $G_2$ flavour symmetry is also broken due to the presence of these massive strings between the D-branes.  These mass terms give rise to the discrete flavour fugacities for the $G_2-E_6$ correspondence between the chiral algebras. In principle, we could have turned on all 6 fugacities but turning on two fugacities gives rise to flavoured $G_2$ characters.  Once we start reducing the gap between D-branes it is natural to expect that we will get unflavoured $G_2$ characters but there is an enhancement of the symmetry and we get the $E_6$ symmetry instead. 

\subsection{Line defects and defect Schur indices}

We showed earlier that the modular data of the subcategory $\mathcal{O}$ of chiral algebra $G_{2,-5/3}$ is related to that of $A_{1,-4/3}$ algebra through a Galois conjugation. The Verlinde fusion algebra of both chiral algebras is the same since the fusion algebra of the highest weight primaries is protected under Galois conjugation.\footnote{Note that although the fusion coefficients are not positive definite, they are like $\mathbb{Z}_3$ fusion rules apart from a $\mathbb{Z}_2$ factor, as we saw above and explained more in \cite{Buican:2019huq}.} In this subsection, we look at some simple applications of the relation discovered in the previous subsection, \ref{galois_conjugation_section}, to predict features of the conjectured 4-dimensional $\mathcal{N}=2$ SCFT with $G_2$ flavour symmetry at level $k_{4D}=10/3$. 

We know that the Schur index of the 4-dimensional $\mathcal{N}=2$ SCFT with a flavour symmetry $G_2$ is the vacuum character of the chiral algebra $G_{2,-5/3}$,
\begin{equation}
    \mathcal{I}_{0}^{G_2}(q)=q^{-c_{4D}/2}\chi_{0}^{G_{2,-5/3}}(q)\, .
\end{equation}
It means that the local operators in the 2-dimensional chiral algebra, which are counted by the vacuum character, are in one-to-one correspondence with the Schur operators counted by the Schur index. In the presence of line operators $L_{a}$, we can similarly write line defect Schur indices, which could be written in terms of characters of other highest weight primaries of the chiral algebra $G_{2,-5/3}$ as expected from calculations for various AD theories, for example in \cite{Cordova:2016uwk}. In the 4-dimensional manifold $\mathbb{R}^{4}$, these full line operators that intersect the chiral plane \cite{Beem:2013sza} at the origin preserve half the supersymmetry \cite{Gaiotto:2010be, Cordova:2016uwk}.
One of the main observations of \cite{Cordova:2016uwk} was to observe that the full defect line Schur indices can be written as,
\begin{equation}\label{schur_index_characters}
    \mathcal{I}_{L_{a,i}}(q)=\sum_{b\in \Lambda} v_{a,i}^{b}(q)\chi_b(q)\, ,
\end{equation}
where $v_{a,i}^{b}$ in general polynomials in of $q,x_1,$ and $x_2$, and the sum is over the highest weight modules whose set is denoted by $\Lambda$. This implies that the indices $b=0,1,2$. With two half-lines inserted,
\begin{equation}
    \mathcal{I}_{L_{a,i}L_{b,j}}(q)=\sum_{b\in \Lambda} v_{a,i,b,j}^{c}(q)\chi_c(q)\, ,
\end{equation}
with another set of polynomials $v_{a,i}^{b}$. In the limit $(q=1)$ these polynomials reduce to,
\begin{equation}
    V_{a}^{c}\equiv v_{a,i}^{c}(q=1)\, ,\nonumber\\
    V_{a,b}^{c}\equiv v_{a,i,b,j}^{c}(q=1)\, ,
\end{equation}
 with the identification $V_{a,0}^{c}=V_a^c$. These coefficients satisfy the fusion algebra,
 \begin{equation}
     V_{a,b}^c=\sum_{a',b'}\mathcal{N}_{a',b'}^{c}V_{a}^{a'}V_{b}^{b'}\, ,
 \end{equation}
where $\mathcal{N}_{a,b}^{c}$ are the Verlinde fusion rules coefficients, 
\begin{equation}
	N_{a,b}^c=\sum_l\frac{S_{al}S_{bl}(S^{-1})_{cl}}{S_{0l}}\notin\mathbb{Z}_{\geq 0}\, .
\end{equation}
These can be derived for the $G_{2,-5/3}$ chiral algebra from the modular-$S$ matrix we found in \eqref{g2_-5_3_s-matrix}. 
Using the Verlinde fusion rules from the modular-$S$ matrix of $G_{2,-5/3}$ and using the fact that $V_{0}^{a}=(1,0,0)$, we can find the remaining coefficients,
\begin{equation}
 V_1^a=(0,-1,1)\, ,\quad V_2^a=(1,-1,1)\, .
\end{equation}
In fact, since Galois conjugation preserves the Verlinde fusion coefficients, the coefficients $V_a^b$ match with the coefficients for $(A_1,A_3)$ AD theory, as found in \cite{Cordova:2016uwk}. 


Thus, starting from the unflavoured $G_2-E_6$ correspondence, we reached the conclusion that since the chiral algebra $A_{1,-4/3}$ of the $(A_1,A_3)$ AD theory is the Galois conjugate of the $G_{2,-5/3}$ chiral algebra, the line defect Schur indices of the $(A_1,A_3)$ AD theory and the conjectured $G_2$ flavour symmetry SCFT should have the same form. In particular, there are three highest weight primaries of $A_{1,-4/3}$ algebra \ref{table: A_1_43_reps}, and associated with each primary are a bunch of defect lines $L_{a,i}$, which are labelled by the primary and an additional index counting degeneracy, so $1\leq a\leq 3$ and the index $1\leq i\leq 3$. The defect Schur indices are \cite{Cordova:2016uwk},
\begin{align}
    \mathcal{I}_{0}^{A_1}(q)&=\chi_0^{A_1}(q)\, ,\nonumber\\
    \mathcal{I}_{1}^{A_1}(q)&=q^{-\frac{1}{2}}\left(-\chi_1^{A_1}(q)+\chi_2^{A_1}(q)\right)\, ,\nonumber\\
    \mathcal{I}_{2}^{A_1}(q)&=q^{-\frac{1}{2}}\left(\chi_0^{A_1}(q)-\chi_1^{A_1}(q)+\chi_2^{A_1}(q)\right)\, .
\end{align}
Since we know the coefficients $V_a^b$, we can write the linear combination of the characters of $G_{2,-5/3}$, which constitutes the line defect Schur indices of the $G_2$ SCFT, up to an overall $q^{-\frac{\alpha}{2}}$, which we have not computed explicitly. Thus, we have the line defect Schur indices of the $G_2$ SCFT,
\begin{align}
    \mathcal{I}_{0}^{G_2}(q)&=\chi_0^{G_2}(q)\, ,\nonumber\\
    \mathcal{I}_{1}^{G_2}(q)&=q^{-\frac{\alpha}{2}}\left(-\chi_1^{G_2}(q)+\chi_2^{G_2}(q)\right)\, ,\nonumber\\
    \mathcal{I}_{2}^{G_2}(q)&=q^{-\frac{\alpha}{2}}\left(\chi_0^{G_2}(q)-\chi_1^{G_2}(q)+\chi_2^{G_2}(q)\right)\, .
\end{align}

\section{Flavouring the DC-MMS Series}\label{flavouredDC}

 We have demonstrated the efficacy of our method to add discrete flavour fugacity in the $A_1/A_2$ correspondence and the $G_2-E_6$ correspondence in the previous section. We follow the same procedure in this section to flavour the $A_{2,-3/2}\mapsto D_{4,1}$ correspondence. We also present the discrete flavour values where the $G_{2,-5/3}\mapsto E_{6,1}$ correspondence holds.

\subsection{\texorpdfstring{$A_{2,-3/2}$}{A2 at level -3/2} to \texorpdfstring{$D_{4,1}$}{D4 at level 1}}
Here we introduce discrete flavour fugacity in the remaining member of the DC-MMS series where we have not seen flavoured correspondence, the $A_{2,-3/2}\mapsto D_{4,1}$ correspondence. The unflavoured correspondence $A_{2,-3/2}\mapsto D_{4,1}$ \cite{Buican:2017rya},
\begin{align}
    \chi_{0}^{A_{2,-3/2}}(q)\sim \chi_{\mathbf{8}_v}^{D_{4,1}}(q)=\chi_{\mathbf{8}_s}^{D_{4,1}}(q)=\chi_{\mathbf{8}_c}^{D_{4,1}}(q)\, ,\nonumber\\
    \chi_{1}^{A_{2,-3/2}}(q)+\chi_{2}^{A_{2,-3/2}}(q)-\chi_{3}^{A_{2,-3/2}}(q)\sim \chi_{\bf 1}^{D_{4,1}}(q)\, ,
\end{align} 
can be expanded to be valid at discrete values of flavour fugacity. To make this statement precise, we will introduce the conformal embedding $A_{2,3}\subset D_{4,1}$. The unflavoured characters $\chi_{1/2}^{D_{4,1}}$ are identified as,
\begin{equation}\label{d4-a2-char-relations}
    \chi_{1}^{D_{4,1}}=\chi_{\bf 1}^{A_{2,3}}+\chi_{\bf 10}^{A_{2,3}}+\chi_{\overline{\mathbf{10}}}^{A_{2,3}}\, ,\quad  \chi_{\mathbf{8}_v}^{D_{4,1}}=\chi_{\mathbf{8}_s}^{D_{4,1}}=\chi_{\mathbf{8}_c}^{D_{4,1}}=\chi_{\bf 8}^{A_{2,3}}.
\end{equation}

The identification above \eqref{d4-a2-char-relations} relates the two partition functions as follows,
\begin{align}
    \mathcal{Z}[D_{4,1}]&=|\chi_{1}^{D_{4,1}}|^2+|\chi_{\mathbf{8}_v}^{D_{4,1}}|^2+|\chi_{\mathbf{8}_s}^{D_{4,1}}|^2+|\chi_{\mathbf{8}_c}^{D_{4,1}}|^2\, ,\nonumber\\
    \mathcal{Z}[A_{2,3}]&=|\chi_{\bf 1}^{A_{2,3}}+\chi_{\bf 10}^{A_{2,3}}+\chi_{\overline{\mathbf{10}}}^{A_{2,3}}|^2+3|\chi_{\bf 8}^{A_{2,3}}|^2\, .
\end{align}

The characters $\chi_{8_v}^{D_{4,1}}$ and $\chi_{\bf 1}^{D_{4,1}}$ can be decomposed into characters of the finite algebra $D_4$ and $A_2$, respectively, at each grade in the energy eigenspace given in tables \ref{table: char_table_d4_level_1} and \ref{table: char_table_A2-3}. We compare the finite character sum and find that we can turn off two fugacities and identify the remaining two fugacities with the fugacities of $A_2$,\footnote{We note that this is one of the relations between the fugacities of $D_4$ and $A_2$ as we note for $E_6$ and $G_2$ in \eqref{eq:transform2}.}
\begin{align}\label{d4-a2-fugacity-rel}
    x_1&=1\, ,\quad x_4=1\, ,\nonumber\\
    x_2&=\frac{1}{y_1y_2}\, ,\quad x_3=\frac{y_2}{y_1^2}\, ,
\end{align}
where the fugacities $x_1,x_2,x_3,x_4$ are $D_4$ fugacities and $y_1,y_2$ are $A_2$ fugacities. Due to the $D_4$ triality, the relation \eqref{d4-a2-fugacity-rel} also holds if we consider the characters of spinor and conjugate spinor representations $\chi_{\mathbf{8}_s}$ and $\chi_{\mathbf{8}_c}$, respectively.

Once we have established the relation between the characters of $A_{2,3}$ RCFT and $D_{4,1}$ RCFT, we proceed to find the discrete flavour fugacity values at which the correspondence $A_{2,-3/2}\mapsto A_{2,3}$ holds, which are the solutions of the equation,
\begin{equation}\label{fugacity_match-ansatz_a2}
    \chi^{A_{2,3}}_{\bf 8}(q,x_1^i,x_2^i)=\mathcal{N}^{A_{2}}(x_1^i,x_2^i)\chi^{A_{2,-3/2}}_{[-3/2;0,0]}(q,x_1^i,x_2^i)\, ,
\end{equation}
to find the discrete set of flavour fugacities. We use the flavoured characters $\chi^{A_{2,-3/2}}_{[-3/2;0,0]}(q,x_1,x_2)$ and $\chi^{A_{2,3}}_{\bf 8}(q,x_1,x_2)$, which are obtained from the branching rules provided in tables \ref{table: char_table_A2-32} and \ref{table: char_table_A2-3}, respectively.
\begin{table}
\centering
\begin{tabular}{ |p{2cm}|p{8cm}|  }
\hline
grade $[q^n]$ & $A_2$ representations and their multiplicities \cite{Buican:2015ina}\\
\hline
0 & {\bf 1} \\
\hline
1 & {\bf 8} \\
\hline
2 & {\bf 1}, {\bf 8}, {\bf 27}\\
\hline
3 & {\bf 1}, 2({\bf 8}), {\bf 10}, {$\overline{\bf 10}$}, {\bf 27}, {\bf 64}\\
\hline
4 & 2({\bf 1}), 4({\bf 8}), {\bf 10}, {$\overline{\bf 10}$}, 3{(\bf 27)}, {\bf 35}, {$\overline{\bf 35}$}, {\bf 64}, {\bf 125}\\
\hline
\end{tabular}
\caption{The multiplicities of each $A_2$ representation which occurs at the first five grades in the representation $[-3/2;0,0]$ of $A_{2,-3/2}$ algebra is tabulated.}
\label{table: char_table_A2-32}
\end{table} 

\begin{table}
\centering
\begin{tabular}{ |p{1.8cm}|p{8.5cm}|  } 
\hline
grade $[q^n]$ & $D_4$ representations and their multiplicities \\
\hline
0 & {$\mathbf{8}_v$} \\
\hline
1 &  {$\mathbf{8}_v$}, {$\mathbf{56}_v$}\\
\hline
2 & 2({$\mathbf{8}_v$}), 2({$\mathbf{56}_v$}), {$\mathbf{160}_v$}\\
\hline
3 & 4({$\mathbf{8}_v$}), 4({$\mathbf{56}_v$}), 2({$\mathbf{160}_v$}), {$\mathbf{224}_{cv}$},  {$\mathbf{224}_{sv}$}\\
\hline
4 & 7({$\mathbf{8}_v$}), 8({$\mathbf{56}_v$}), 5({$\mathbf{160}_v$}), 2({$\mathbf{224}_{cv}$}), 2({$\mathbf{224}_{sv}$}),\\ & {$\mathbf{840}_v$}, {$\mathbf{112}_v$}\\
\hline
\end{tabular}
\caption{The multiplicities of each $D_4$ representation which occurs at the first five grades in the representation $[0;1,0,0,0]$ of $D_{4,1}$ algebra is tabulated.}
\label{table: char_table_d4_level_1}
\end{table}

\begin{table}
\centering
\begin{tabular}{ |p{1.8cm}|p{8.5cm}|  } 
\hline
grade $[q^n]$ & $A_2$ representations and their multiplicities \cite{kass1990affine} \\
\hline
0 & {\bf 8} \\
\hline
1 &  {\bf 1}, 2({\bf 8}), {\bf 10}, {$\overline{\bf 10}$}, {\bf 27}\\
\hline
2 & 2({\bf 1}), 6({\bf 8}), 3({\bf 10}), 3({$\overline{\bf 10}$}), 4({\bf 27}), {\bf 35}, {$\overline{\bf 35}$}\\
\hline
3 & 4({\bf 1}), 16({\bf 8}), 8({\bf 10}), 8({$\overline{\bf 10}$}), 12({\bf 27}), 4({\bf 35}), 4({$\overline{\bf 35}$}), 2({\bf 64})\\
\hline
4 & 10({\bf 1}), 36({\bf 8}), 21({\bf 10}), 21({$\overline{\bf 10}$}), 32({\bf 27}), {\bf 28}, {$\overline{\bf 28}$}, 12({\bf 35}), 12({$\overline{\bf 35}$}), 8({\bf 64}), {\bf 81}, {$\overline{\bf 81}$}\\
\hline
\end{tabular}
\caption{The multiplicities of each $A_2$ representation which occurs at the first five grades in the representation $[1;1,1]$ of $A_{2,3}$ algebra is tabulated.}
\label{table: char_table_A2-3}
\end{table}
The complete set of discrete fugacities satisfying eq \ref{fugacity_match-ansatz_a2} are 34 in number. Many discrete fugacity values can be obtained from a set of \textit{independent fugacities} by applying the two symmetry transformations we explain next.

The first symmetry is an exchange symmetry, i.e., for the i-th solution $(x_1^i,x_2^i)$, there exists another solution $(x_2^i,x_1^i)$,
   \begin{equation}\label{symmetry_1}
        (x_1^i,x_2^i)\to (x_2^i,x_1^i)\, .
    \end{equation} 
The second symmetry is the complex conjugation, i.e., for the i-th solution $(x_1^i,x_2^i)$, there exists another solution $(\bar x_1^i,\bar x_2^i)$,
    \begin{equation}\label{symmetry_2}
        (x_1^i,x_2^i)\to (\bar x_1^i,\bar x_2^i)\, .
    \end{equation}
In fact, as we will see, these symmetries exist for the solutions of $G_2$ ansatz \eqref{fugacity_match-ansatz_g2} as well.
In the set of 34 solutions, we have 13 solutions with a distinct pair under symmetry \eqref{symmetry_1}. The remaining 8 solutions are symmetric but not distinct. Thus, \eqref{symmetry_1} reduces the number from 34 to 21. There is a solution $(e^{-2\pi i/3},e^{2\pi i/3})$ for which \eqref{symmetry_1}, \eqref{symmetry_2} are the same symmetries. Thus, complex conjugation symmetry \eqref{symmetry_2} has 10 solutions, with a complex conjugate in the remaining 21 solutions. Including the solution $(1,1)$ and the solution $(e^{-2\pi i/3},e^{2\pi i/3})$, we have a total of 12 independent fugacities. In the following, we again use the notation $\xi(n)=e^{2\pi i/n}$ to denote the $n$-th root of unity. The 12 independent solutions are listed below,
\begin{align}\label{dep_a2_flavour_fugacities}
    \begin{pmatrix}
        x_1^i\\
        x_2^i
    \end{pmatrix}=
    \left(
    \begin{array}{cccccccccccc}
        1 & \xi(7)^1 & \xi(7)^1 & \xi(7)^2 & \xi(9)^1 & \xi(9)^1 & \xi(9)^1 & \xi(9)^2 & \xi(9)^2 & \xi(9)^3 & \xi(9)^3 & \xi(9)^4 \\
        1 & \xi(7)^2 & \xi(7)^4 & \xi(7)^8 & \xi(9)^1 & \xi(9)^4 & \xi(9)^7 & \xi(9)^2 & \xi(9)^5 & \xi(9)^3 & \xi(9)^6 & \xi(9)^4
     \end{array}
     \right)\, .
\end{align}
The solutions occur at only the 7th and 9th roots of unity. Note that the fugacity $(y_1=-1,y_2=-1)$ is a solution of the eqn \eqref{fugacity_match-ansatz_a2} but the left hand and the right hand side vanish, although this solution is a non-trivial solution for \eqref{fugacity_match-ansatz_g2}.

\subsection{\texorpdfstring{$G_{2,-5/3}$}{G2 at level -5/3} to \texorpdfstring{$E_{6,1}$}{E6 at level 1}}\label{explicit_fug_g2}
We now move to the explicit set of solutions to the equation \eqref{fugacity_match-ansatz_g2} we promised earlier. In section \ref{Flavouring G2/E6 correspondence}, we have shown the relation \eqref{e6-flavours-to-g2-flavours} between the flavour fugacity of $E_6$ and $G_2$. The solutions of the eqn \eqref{fugacity_match-ansatz_g2},
\begin{equation}\label{fugacity_match-ansatz_g2sol}
        \chi^{G_{2,3}}_{\bf 27}(q,y_1^i,y_2^i)=\mathcal{N}(y_1^i,y_2^i)\chi^{G_{2,-5/3}}_{0}(q,y_1^i,y_2^i)\, ,
    \end{equation}
up to the above two symmetries, \eqref{symmetry_1} and \eqref{symmetry_2} are just 17. The counting can be explained as follows, the 62 fugacities have a solution $(y_1=-1,y_2=-1)\, ,(y_1=1,y_2=1)$, apart from which the fugacities occur in pairs due to \eqref{symmetry_1} and another pair due to \eqref{symmetry_2}. The number of independent fugacities is reduced from 60 solutions to $60/4=15$. Thus, the total number of solutions is 17. We use $\xi(n)=e^{2\pi i/n}$ to denote that the set of 17 independent roots can be represented as,

\begin{align}
(y_1^i,y_2^i)=\left(
\begin{array}{cc}
 1\hspace{.8 cm} ,& 1 \\
 \xi(2)\, \, \, \, \, , & \xi(2) \\
 \xi(7)^1\, \, \, \,  , & \xi(7)^2 \\
 \xi(7)^1\, \, \, \,  , & \xi(7)^4 \\
 \xi(7)^2\, \, \, \, , & \xi(7)^4 \\
 \xi(10)^1\, , & \xi(10)^2 \\
\xi(10)^1\, , & \xi(10)^3 \\
 \xi(10)^1\, , & \xi(10)^6 \\
 \xi(10)^1\, , & \xi(10)^7 \\
 \xi(10)^2\, , & \xi(10)^7 \\
 \xi(10)^4\, , & \xi(10)^7 \\
 \xi(13)^1\, , & \xi(13)^3 \\
\xi(13)^1\, , & \xi(13)^9 \\
 \xi(13)^2\, , & \xi(13)^5 \\
 \xi(13)^2\, , & \xi(13)^6 \\
 \xi(13)^3\, , & \xi(13)^9 \\
 \xi(13)^5\, , & \xi(13)^6
\end{array}
\right)\, ,\quad i=\{1,\cdots,17\}\, .
\end{align}\label{discrete fugacity G2}

The significance of the specific discrete flavour fugacities at the 7th, 10th, and 13th roots of unity is not completely understood by us, and we hope to answer this in a future work.

\section{Conclusions}\label{conclusions}

We have explored the relation between the flavoured characters of the highest weight representations in category $\mathcal{O}$ of the fractional-level current algebra $\log$ CFTs in the DC series with the level one current algebra RCFTs in the MMS series \eqref{non-unitary to unitary series}. In particular, we have presented a complete set of \textit{unitarisation} maps between the unflavoured characters in the \textit{even sector} of the DC series and MMS series. We then proceeded to add discrete flavour fugacity in the correspondence in a canonical way discussed in detail in the main text following \cite{Mukhi:1989bp, Buican:2019huq}. We identify two symmetries, \eqref{symmetry_1} and \eqref{symmetry_2}, which allow us to consider only a smaller set of \textit{independent fugacities} in the case of $A_{2,-3/2}\mapsto A_{2,3}$ and $G_{2,-5/3}\mapsto G_{2,3}$ flavoured correspondence.

We have paid special attention to the correspondence between $G_{2,-5/3}$ and the $E_{6,1}$ CFTs in our bid to propose an explicit F-theory construction of the four-dimensional $\mathcal{N}=2$ SCFT with a $G_2$ flavour symmetry. We have used a two-pronged approach. Firstly, the unflavoured $G_2-E_6$ correspondence is flavoured via the conformal embedding $G_{2,3}\subset E_{6,1}$, as explained in section \ref{Flavouring G2/E6 correspondence}. Secondly, we have shown a relation between the category $\mathcal{O}$ modular data of $A_{1,-4/3}$ and $G_{2,-5/3}$ $\log$ CFTs via a Galois conjugation action, which hints towards the existence of the Higgs branch of the 4D unitary SCFT with a $G_2$ flavour symmetry. 
Through the Coulomb branch geometry, the rank 1 $\mathcal{N}=2$ 4D SCFTs was characterised and identified in \cite{Argyres:2015ffa, Argyres:2015gha, Argyres:2016xmc, Caorsi:2018zsq}) whose associated Magnetic quiver description was provided in \cite{Bourget:2020asf}. 
The F-theory construction of the 4D SCFT with $E_6$ flavour symmetry is well-known following the results of \cite{Dasgupta:1996ij}. Since we have found the characters of category $\mathcal{O}$ of the $G_{2,-5/3}$ chiral algebra starting from the chiral algebra $E_{6,1}$, albeit at discrete flavour fugacities, we propose that the Higgs branch of the SCFT with a broken $G_2$ flavour symmetry is a subbranch of the $E_6$ Higgs branch. From an F-theory point of view, we know that the Higgs branch of rank one theories has only ADE-type singularities. Thus, to construct the 4D unitary SCFT with $G_2$ flavour symmetry, we need to look at the $E_6$ singularity point $\tau=e^{\pi i/3}$ in the moduli space of F-theory coupling, turn off two fugacities, and identify the remaining four as given in eq.\eqref{e6-flavours-to-g2-flavours}. It would be interesting to get to an explicit construction of the $G_2$ Higgs branch from the F-theory construction. In particular, it would be nice to find out the geometric interpretation of turning off some $E_6$ fugacities to access the $G_2$ branch. This presumably corresponds to the asymmetric approach of 7-branes to form the $E_6$ singularity.

The Galois conjugation relation between the category $\mathcal{O}$ modular data of $A_{1,-4/3}$ and the $G_{2,-5/3}$ CFTs is a powerful tool that should allow computations of observables in the $G_2$ SCFT directly from the well-understood $(A_1,A_3)$ AD theory. For example, previously in \cite{Buican:2019huq} Galois conjugation relations were used  to compute data of the chiral algebra of the 4D infrared theory from the chiral algebra of the ultraviolet theory. We demonstrate that the defect Schur indices of the $G_2$ SCFT are the same as the defect Schur indices of the $(A_1, A_3)$ AD theory, up to an overall $q$ factor, by utilising the Galois conjugation relation between their respective chiral algebras.

We have observed that the $F_{4,-5/2}$ chiral algebra contains a quasi-character in its even sector. The Schur index of the associated 4D SCFT is a vector-valued modular form which will contain a quasi-character. Interestingly, the 4D SCFT corresponding to the chiral algebra $F_{4,-5/2}$ is found to be anomalous \cite{Shimizu:2017kzs}. It would be interesting to relate these two statements in the future.

Although originally in the context of rank 1 theories, the relations between the 2D chiral algebras of different 4D SCFTs can be extended to higher-rank theories. The problem of classification quickly gets complicated partly because the 2D CFTs for 4D rank $r$ theories would minimally be $r+1$ character theories. The classification of 2D CFTs with a fixed number of characters is a notoriously difficult problem, and it has only recently been solved for 2-character theories with $c<25$ \cite{Mukhi:2022bte}. It is entirely possible that the sector of three and higher character theories relevant for the 4D classification may simply be a subsector of the 2D CFT classification problem and, therefore, may still be under control from the 4D theory point of view.

\vspace{0.5cm}
 \subsection*{Acknowledgement}

We thank Antoine Bourget, Abhijit Gadde, Madalena Lemos, Sunil Mukhi, Yiwen Pan, and Ashoke Sen for helpful discussions and comments. We also thank the anonymous referee for valuable comments which helped improve our manuscript. We thank the organisers of \textit{Indian Strings Meeting 2023} at the Indian Institute of Technology Bombay for facilitating discussions. MA, SG, and KS acknowledge the Infosys foundation for their generous support. MA thanks the Harish-Chandra Research Institute, Allahabad (HRI) and the Institute of Mathematical Sciences, Chennai for hospitality. KS would like to thank HRI, and the Chennai Mathematical Institute for hospitality.

 \appendix

\section{Notations and conventions}\label{Notations}


A highest weight $\lambda$ of an affine algebra $\hat{\mathfrak{g}}$  associated with a finite Lie algebra $\hat{\mathfrak{g}}$ of rank $r$ is represented in terms of its Dynkin labels $[\lambda_0;\lambda_1, \lambda_2,\cdots, \lambda_r]$. An integrable representation has Dynkin labels $\lambda\in\mathbb{Z}_{\geq 0}$. The level of the affine algebra is,
\begin{equation}
    k=\sum_{i=0}^{r} a_{i}^\vee \lambda_i\, ,
\end{equation}
where $a_{i}^\vee$ are the comarks of the affine algebra $\hat{\mathfrak{g}}$ with $a_{0}^\vee=1$ in the standard convention. The character of an integrable representation $[\lambda_0;\lambda_1, \lambda_2,\cdots, \lambda_r]$ of $\hat{\mathfrak{g}}$ can be written in terms of the characters of $\mathfrak{g}$ of weights in the weight system $\Lambda$ of the highest weight $[\lambda_1, \lambda_2,\cdots, \lambda_r]$. The characters of $\mathfrak{g}$ are defined as,
\begin{equation}\label{general_def_char}
 \mathrm{ch}(x_1,x_2,\cdots,x_r)=\sum_{\rho\in\Lambda} x_1^{\rho_1}x_2^{\rho_2}\cdots x_r^{\rho_r}\, ,
\end{equation}
where the sum is performed over the weights $\rho=[\rho_1, \rho_2,\cdots, \rho_r]\in\Lambda$ and $x_i\, \text{with, }\, 1\leq i\leq r$ are the fugacities. Note that there is no preferred basis for expanding the weight $\lambda$ in terms of its Dynkin labels, and we will make a different choice based on the simplification of the results.

\subsection{Character formula for \ensuremath{A_1} and \ensuremath{A_2}}\label{a2 basics}

The character corresponding to the integrable spin $s$, i.e. $2s\in\mathbb{Z}_{\geq 0}$ and with dimension of the irrep. $\mathrm{dim}\mathcal{R}=2s+1$ of the finite Lie algebra $A_1$ can be written in a compact form,
\begin{equation}
    \mathrm{ch}_{\mathcal{R}}^{A_1}(x)=\frac{x^{\mathcal{R}}-x^{-\mathcal{R}}}{x-x^{-1}},
\end{equation}
where $x$ is the fugacity corresponding to the Cartan of $A_1$.

The character formula for finite Lie algebra $A_2$ in an irreducible representation $\mathcal{R}$ with the Dynkin labels $[\lambda_1,\lambda_2]$ can be written in terms of Vandermonde determinant as,
\begin{align}
\mathrm{ch}_{\mathcal{R}}^{A_2}=\frac{det(w_i^{l_j+3-j})}{det(w_i^{3-j})},\quad \forall \, i,j\in\{1,2,3\},
\end{align}
where, $w_1=x_1$, $w_2=x_2^{-1}$, and $w_3=x_2x_1^{-1}$ with $l_1=\lambda_1+\lambda_2$, $l_2=\lambda_2$, and $l_3=0$ respectively.

\subsection{Elements of \ensuremath{G_2}}\label{g2 basics}

Exceptional group $G_2$ has rank $2$ with the following Dynkin diagram, 
\begin{equation*}
	    \begin{tikzpicture}                             
		\draw (-0.5, 0) -- (0.5,0);
		\draw (-0.5, 0.1) -- (0.5,0.1);
        \draw (-0.5, -0.1) -- (0.5,-0.1);
		\node[circle,fill=black,inner sep=1.5mm] at (-0.5,0) {};
        \node[circle,fill=black,inner sep=1.5mm] at (0.5,0) {};
		\node at (0, 0) {$\big{<}$};
        \node at (-0.5, -0.4) {$(1;2)$};
        \node at (0.5, -0.4) {$(2;1)$};
	\end{tikzpicture}
 \end{equation*}
 where the $i$-th node is denoted by $(i,a_i^\vee)$, and $a_i^\vee$ are the corresponding co-marks. The Cartan matrix of $G_2$,
 \begin{equation}
    A_{G_2}=\begin{pmatrix}
2 & -1\\
-3 & 2
\end{pmatrix}
 \end{equation}
 and the set of positive roots, with $\alpha_1$ to be the short root and $\alpha_2$ to be the longer one,
 \begin{equation}
     {\Delta}_+\equiv \{\alpha_1,\alpha_2,\alpha_1+\alpha_2,2\alpha_1+\alpha_2,3\alpha_1+\alpha_2,3\alpha_1+2\alpha_2\}.
 \end{equation}
 The fundamental weights of the $G_2$ group are,
 \begin{equation}
     \omega_1=2\alpha_1+\alpha_2, \qquad \mathrm{and} \qquad \omega_2=3\alpha_1+2\alpha_2.
 \end{equation}
In this convention, the root diagram of $G_2$ algebra is,
 
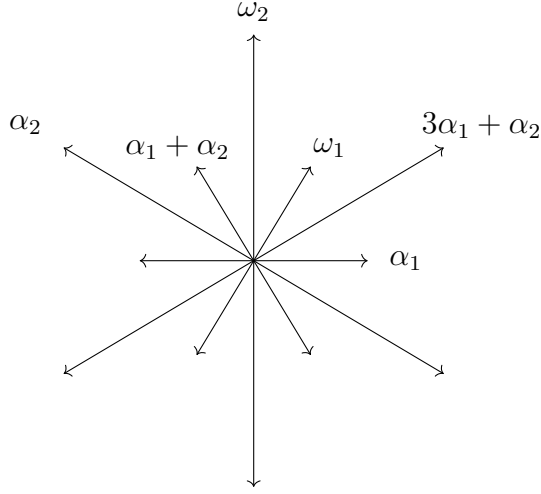
\begin{equation*}
	    \begin{tikzpicture}                             
		\draw[->] (0, 0) -- (-0.75,1.25);
		\draw[->] (0, 0) -- (0.75,1.25);
        \draw[->] (0, 0) -- (-0.75,-1.25);
		\draw[->] (0, 0) -- (0.75,-1.25);
        \draw[->] (0, 0) -- (-1.5,0);
        \draw[->] (0, 0) -- (1.5,0);
        \draw[->] (0, 0) -- (0,3);
        \draw[->] (0, 0) -- (0,-3);
        \draw[->] (0, 0) -- (2.5,-1.5);
        \draw[->] (0, 0) -- (2.5,1.5);
        \draw[->] (0, 0) -- (-2.5,1.5);
        \draw[->] (0, 0) -- (-2.5,-1.5);
        \node at (2, 0) {$\alpha_1$};
        \node at (-3, 1.8) {$\alpha_2$};
        \node at (3, 1.8) {$3\alpha_1+\alpha_2$};
        \node at (-1, 1.5) {$\alpha_1+\alpha_2$};
        \node at (1, 1.5) {$\omega_1$};
        \node at (0, 3.3) {$\omega_2$};
	\end{tikzpicture}
 \end{equation*}
The $G_2$ group has two fundamental representations of dimensions $7$ and $14$, respectively. Its adjoint representation is $14$ dimensional, and the dihedral group $D_6$ of order $12$ is the Weyl group of $G_2$, $W(G_2)=D_6$. Following the notation of \cite{crew2018branching}, we can write the finite dimensional character of $G_2$ with respect to the Dynkin index $[k,l]:=k\omega_1+l\omega_2$ as,
\begin{equation}\label{finite character of G2}
    \mathrm{ch}_{[k,l]}^{(G_2)}(y_1,y_2)=\frac{N_{k,l}}{D},
\end{equation}
where,
\begin{align}
    N_{k,l} = & y_1^{k + 2 l + 3} y_2^{k + l + 2} + y_3^{k + 2 l + 3} y_1^{k + l + 2} + y_2^{k + 2 l + 3} y_3^{k + l + 2} + y_1^{-(k + 2 l + 3)} y_2^{-(k + l + 2)} \nonumber\\ 
   & + y_2^{-(k + 2 l + 3)} y_3^{-(k + l + 2)} + y_3^{-(k + 2 l + 3)} y_1^{-(k + l + 2)} - y_1^{k + 2 l + 3} y_3^{k + l + 2} - y_3^{k + 2 l + 3} y_2^{k + l + 2} \nonumber\\
   & - y_2^{k + 2 l + 3} y_1^{k + l + 2} - y_1^{-(k + 2 l + 3)} y_3^{-(k + l + 2)}-  y_3^{-(k + 2 l + 3)} y_2^{-(k + l + 2)} - y_2^{-(k + 2 l + 3) }y_1^{-(k + l + 2)}\nonumber\\
   D = & (y_1-y_2)(y_1-y_3)(y_2-y_3)(1-y_1)(1-y_2)(1-y_3).
\end{align}
In this notation, $y_i=e^{L_i}$ and $y_1y_2y_3=1$. We can express the positive roots in terms of $L_1$ and $L_2$ as follows,
\begin{align}\label{positive root of G2}
    {\Delta}_+ & \equiv \{L_2,(L_1-L_2),L_1,(L_1+L_2),(L_1+2L_2),(2L_1+L_2)\}\nonumber\\
    & \equiv \{\alpha_1,\alpha_2,\alpha_1+\alpha_2,2\alpha_1+\alpha_2,3\alpha_1+\alpha_2,3\alpha_1+2\alpha_2\}.
\end{align}
With this above mapping we have,
\begin{equation}\label{G2 roots at y1 y2}
    e^{\alpha_1}:=y_2, \qquad \text{and} \qquad e^{\alpha_2}:= \frac{y_1}{y_2}.
    \end{equation}
For example the $27$ dimensional irreducible representation (irrep.) of $G_2$ has the following finite character,
\begin{align}\label{27 G2}
    \mathrm{ch}_{\bf{27}}^{G_2}=\mathrm{ch}_{[0,2]}^{G_2}=& 3+\frac{1}{y_1^2}+\frac{2}{y_1}+y_1^2+2y_1+\frac{1}{y_2^2}+\frac{2}{y_2}+y_2^2+2y_2\nonumber\\
    &+\frac{y_1}{y_2}+\frac{y_2}{y_1}+2y_1y_2+\frac{2}{y_1y_2}+y_1y_2^2+y_1^2y_2\nonumber\\
    &+\frac{1}{y_1y_2^2}+\frac{1}{y_1^2y_2}+y_1^2y_2^2+\frac{1}{y_1^2y_2^2}.
\end{align}
\subsection{Elements of \ensuremath{E_6}}

We have used the following convention of the Dynkin diagram of $E_6$,
\begin{equation}
\begin{tikzpicture}[baseline={(current bounding box.center)}, scale = 1]
		\tikzstyle{vertex}=[circle, fill=black, minimum size=2pt,inner sep=2pt];
		\def\r{1.2};
		\node[vertex] (T1) at (\r*1,\r*0) {};
		\node[vertex] (T2) at (\r*2,\r*0) {};
		\node[vertex] (T3) at (\r*3,\r*0) {};
		\node[vertex] (T4) at (\r*4,\r*0) {};
		\node[vertex] (T5) at (\r*5,\r*0) {};
		\node[vertex] (T6) at (\r*3,\r*1) {};
		
		\draw[-] (T1) -- (T2);
		\draw[-] (T2) -- (T3);
		\draw[-] (T3) -- (T4);
		\draw[-] (T4) -- (T5);
		\draw[-] (T3) -- (T6);

		\draw[above] (T1) node {$(1;1)$};
		\draw[above] (T2) node {$(2;2)$};
		\draw[below] (T3) node {$(3;3)$};
		\draw[above] (T4) node {$(4;2)$};
		\draw[above] (T5) node {$(5;1)$};
		\draw[above] (T6) node {$(6;2)$};
	\end{tikzpicture}
\end{equation}
with the following Cartan matrix,
\begin{equation}
    A_{E_6}=\begin{pmatrix}
        2 & -1 & 0& 0& 0& 0\\
        -1 & 2 & -1 & 0 & 0 & 0\\
        0 & -1 & 2 & -1 & 0 & -1\\
        0 & 0 & -1 & 2 & -1 & 0\\
        0 & 0 & 0 & -1 & 2 & 0\\
        0 & 0 & -1 & 0 & 0 & 2\\
    \end{pmatrix}\, .
\end{equation}
The rank $r=6$ exceptional group $E_6$ has dimension $78$. Also, it has two fundamental representations with dimension $27$, which are complex conjugate to each other and are represented by $\bf{27}$ and $\bf{\overline{27}}$. For example, the character for the irrep. $\bf{27}$ is given by,
\begin{align}\label{27E6}
  \mathrm{ch}_{\bf{27}}^{E_6}=& \frac{x_5 x_1}{x_2}+\frac{x_6 x_1}{x_3}+\frac{x_3 x_1}{x_2 x_4}+\frac{x_4 x_1}{x_2 x_5}+\frac{x_1}{x_6}+x_1+\frac{x_3}{x_2}+\frac{x_4}{x_3}+\frac{x_2 x_5}{x_3}+\frac{x_5}{x_4}\nonumber\\
  &+\frac{x_4 x_6}{x_3}+\frac{x_5 x_6}{x_4}+\frac{x_6}{x_2}+\frac{x_2 x_6}{x_1 x_3}+\frac{x_6}{x_5}+\frac{x_2}{x_4}+\frac{x_3}{x_1 x_4}+\frac{x_4}{x_1 x_5}+\frac{x_2 x_4}{x_3 x_5}\nonumber\\
  &+\frac{1}{x_5}+\frac{x_4}{x_6}+\frac{x_2}{x_1 x_6}+\frac{x_3}{x_2 x_6}+\frac{x_3 x_5}{x_4 x_6}+\frac{x_3}{x_5 x_6}+\frac{x_2}{x_1}+\frac{x_5}{x_1}\,.
\end{align}
\subsection{Fugacity matching in \ensuremath{G_2} and \ensuremath{E_6}}\label{discrete_flavours}
In section \ref{Flavouring G2/E6 correspondence}, the flavouring correspondence between characters of $E_{6,1}$ and $G_{2,3}$ CFTs is mentioned. From the correspondence, we know that the non-vacuum character $\chi_{\bf{27}}^{E_{6,1}}(q,\mathbf{x})$ of the $E_6$ theory at level $1$ and the character $\chi_{\bf{27}}^{G_{2,3}}(q,\mathbf{y})$ of the $G_2$ at level $3$ theory are related by conformal embedding. To show the flavour fugacity identification at each order of $q^n$, we have the following choices,
\begin{align}\label{eq:transform2}
 & x_1\to 1\,, x_5\to 1\,, \quad\text{and}\quad x_3\to \frac{1}{x_2 x_4}\,, \left(x_6\to \frac{1}{x_2}\,\text{or}\,x_6\to \frac{1}{x_4}\right), \quad \text{with  } x_2\to y_1,\, x_4\to y_2,\nonumber\\
  & x_1\to 1\,, x_5\to 1\, , \quad\text{and}\quad x_3\to \frac{1}{x_2 x_6}\,,\left( x_4\to \frac{1}{x_6}\,\text{or}\,x_4\to \frac{1}{x_2}\right),\quad \text{with  } x_2\to y_1,\, x_6\to y_2,\nonumber\\
 & x_1\to 1\,, x_5\to 1\,, \quad\text{and}\quad x_3\to \frac{1}{x_4 x_6}\,,\left( x_2\to \frac{1}{x_4}\,\text{or}\,x_2\to \frac{1}{x_6}\right),\quad \text{with  } x_4\to y_1,\, x_6\to y_2.\nonumber\\
\end{align}
We can verify the above identification from eqns \eqref{27 G2}, and \eqref{27E6} respectively. The above identification \eqref{eq:transform2} remains true for relating the finite characters of $G_2$ and $E_6$ given in tables \ref{table: g2-level-3-reps} and \ref{table: e6-level-1-reps} respectively.

\section{Vacuum character of \ensuremath{G_{2,-5/3}}}\label{Details of vacuum character}

In this appendix, we give a short review of the Ka\v c-Wakimoto character formula \cite{kac_1990} for the affine Lie algebra at negative fractional level $k$ and calculate the vacuum character for the $G_{2,-5/3}$ chiral algebra. The character of the Verma module $M(\lambda)$ corresponding to the highest weight $\lambda$ of the affine Lie algebra $\hat{\mathfrak{g}}$ is,
\begin{align}
  \text{ch M($\lambda$)} &=e^{\lambda}\prod_{n=1}^\infty\Bigg[\frac{1}{(1-q^n)^{{r}}}\prod_{{\alpha}\in {\Delta}_+}\frac{1}{(1-q^ne^{{\alpha}})(1-q^{n-1}e^{-{\alpha}})}\Bigg],
\end{align}
where $r$ is the rank of the finite Lie algebra $\mathfrak{g}$ with the set of positive roots $\Delta_+=\{\alpha\}$. We define $q=e^{-\delta}$, with $\delta=\sum_{i=0}^r a_i\alpha_i$, where $a_i$  and $\alpha_i$ are the marks and simple roots of $\mathfrak{g}$, respectively. Let us define $\hat{\alpha}_i$ as the simple roots inside the set of real positive roots $\hat{\Delta}^{re}_+$ of $\hat{\mathfrak{g}}$. 

For general affine highest weight $\lambda$, the Verma module $M(\lambda)$ is not always reducible. If the affine weight $\lambda$ is not dominant or integrable, then $W_\lambda$ is the Weyl group associated with the simple Weyl reflections of $\hat{\alpha}_i$. The Weyl reflection of $\lambda$ generated by an element $\omega\in W_\lambda$ is defined by $\omega\cdot\lambda\equiv\omega(\lambda+\rho)-\rho$ with the Weyl vector $\rho$ of $\hat{\mathfrak{g}}$. So, the Ka\v c-Wakimoto character formula for the irreducible module $L(\lambda)$ associated with the Ka\v c-Wakimoto admissible highest weight $\lambda$ is given by,
\begin{align}
   \text{ch L($\lambda$)} &=\sum_{\omega\in W_\lambda} \epsilon(\omega) M(\omega\cdot\lambda).
\end{align}
If $\omega$ is composed of $k$ simple Weyl reflections, then $\epsilon(\omega)=(-1)^k$.

The Ka\v c-Wakimoto character corresponding to the admissible highest weight $[-\frac{5}{3};0,0]$, namely the vacuum character of the chiral algebra $G_{2,-5/3}$, is given by,
\begin{align}
   \text{ch L($\lambda$)}& =\left(\frac{y_1^6y_2^5\left(1-q^2\text{ch}_{27}+q^3\text{ch}_{64}-q^6\text{ch}_{286}+q^8\text{ch}_{729}+{\it O}[q^{12}]\right)}{(y_1-1)(y_2-1)(y_1-y_2)(y_1y_2-1)(y_1^2y_2-1)(y_1y_2^2-1)}\right)\nonumber\\
    & \hspace{2cm}\times\prod_{n=1}^\infty\Bigg[\frac{1}{(1-q^n)^{{2}}}\prod_{{\alpha}\in {\Delta}_+}\frac{1}{(1-q^ne^{{\alpha}})(1-q^{n-1}e^{-{\alpha}})}\Bigg],
\end{align}
where the finite characters $\text{ch}_{\mathcal{R}}$ of irreps. of $G_2$ in terms of fugacities $(y_1,y_2)$  are given in \eqref{finite character of G2}. We used \eqref{positive root of G2} and \eqref{G2 roots at y1 y2} for the set of positive roots of $G_2$ and their exponentiation in terms of $(y_1,y_2)$.

\section{Non-vacuum character of \texorpdfstring{$G_{2,3}$}{G2 at level 3}}\label{non vac G2 3}

The non-vacuum character of the $G_{2,3}$ can be written in terms of the characters of the finite group $G_2$ \eqref{g2-level3-non-vacuum}. Here we discuss the representations of $G_2$ that occur in \eqref{g2-level3-non-vacuum} by relating this character to the vacuum character of $G_{2,-5/3}$.
Fusions between the irreducible representations of $G_2$ are the following,
\begin{align}\label{decompose G2 }
    &\bf{27\otimes 1=27},\nonumber\\
    &\bf{27\otimes 14=7\oplus 14\oplus 27 \oplus 64 \oplus 77 \oplus 189},\nonumber\\
    &\bf{27\otimes 77'=27 \oplus 64 \oplus 77' \oplus 77 \oplus 182\oplus 189\oplus 286\oplus 448\oplus {729}},\nonumber\\
    &{\bf{27\otimes 27=1\oplus 7\oplus 14\oplus 2(27) \oplus 2(64) \oplus 77' \oplus 77 \oplus 182\oplus 189}}.
\end{align}
We write the relation between the characters of ${G_{2,-5/3}}$ and ${G_{2,3}}$ again for convenience,
\begin{equation}\label{G2 compare}
    \chi^{G_{2,3}}_{\bf 27}(q,y_1^i,y_2^i)=\text{ch}_{\bf{27}}^{G_2}(y_1^i,y_2^i)\chi^{G_{2,-5/3}}_{0}(q,y_1^i,y_2^i)\, 
\end{equation}
We compare the left and the right hand sides of \eqref{G2 compare} order by order in $O(q^n)$, see tables \ref{table: g2-m5/3-mutiplicity} and \ref{table: g2-level-3-reps}. Using \eqref{decompose G2 }, we can see upto $O(q)$ both the character decomposition matches in \eqref{G2 compare}. The mismatch starts from $O(q^2)$ onwards and translates that \eqref{G2 compare} is satisfied for a finite set of fugacity values mentioned in \eqref{discrete fugacity G2} instead of every $(y_1,y_2)$. At $O(q^2)$, $\it{rhs}$ of \eqref{G2 compare} gives,
\begin{align}
    \bf{27\otimes(1\oplus 14\oplus 77')}=&\bf{7\oplus 14\oplus 3(27) \oplus 2(64) \oplus 77' \oplus 2(77) \oplus 182\oplus 2(189)}\nonumber\\
    &\bf{\oplus 286\oplus 448\oplus {\cancel{\bf 729}} }.
\end{align}
But in contrast we have from $\it{lhs}$,
\begin{align}
    &\bf{7\oplus 14\oplus 3(27) \oplus 2(64) \oplus 77' \oplus 2(77) \oplus 182\oplus 2(189)\oplus 286\oplus 448\oplus}{ ({\bf{27\otimes27}}) },
\end{align}
which shows that extra contributions are due to $\bf{27\otimes 27}$ instead of the irreducible representation $\bf{729}$. 


 \bibliography{references-2}
\end{document}